\documentclass[12pt]{article}

\usepackage{amsmath,amsfonts,amssymb}
\usepackage{graphicx,cite,epsfig}
\def\slashchar#1{\setbox0=\hbox{$#1$}           
   \dimen0=\wd0                                 
   \setbox1=\hbox{/} \dimen1=\wd1               
   \ifdim\dimen0>\dimen1                        
      \rlap{\hbox to \dimen0{\hfil/\hfil}}      
      #1                                        
   \else                                        
      \rlap{\hbox to \dimen1{\hfil$#1$\hfil}}   
      /                                         
   \fi}  
\def\mx{M_X}
\def\mjjtautau{M_{jj\tauptaum}}

\def\wp{W^+}
\def\wm{W^-}
\def\wpm{W^{\pm}}
\def\mz{m_Z}
\def\epem{e^+e^-}

\def\fbi{~{\rm fb}^{-1}}

\def\pb{~{\rm pb}}
\def\tev{~{\rm TeV}}
\def\gev{~{\rm GeV}}
\def\bit{\begin{itemize}}
\def\eit{\end{itemize}}
\def\ben{\begin{enumerate}}
\def\een{\end{enumerate}}
\def\bed{\begin{description}}
\def\eed{\end{description}}
\def\susy{{\sc Susy}}

\def\msusy{M_{\rm SUSY}}
\def\mt{m_t}
\def\mb{m_b}
\def\mtau{m_\tau}
\def\eps{\epsilon}
\def\lam{\lambda}
\def\tanb{\tan\beta}
\def\mueff{\mu_{\rm eff}}

\def\half{\frac{1}{2}\,}

\def\R{ {\rm R \kern -.31cm I \kern .15cm}}
\def\C{ {\rm C \kern -.15cm \vrule width.5pt \kern .12cm}}
\def\Z{ {\rm Z \kern -.27cm \angle \kern .02cm}}
\def\N{ {\rm N \kern -.26cm \vrule width.4pt \kern .10cm}}
\def\1{{\rm 1\mskip-4.5mu l} }
\def\lsim{\raise0.3ex\hbox{$<$\kern-0.75em\raise-1.1ex\hbox{$\sim$}}}
\def\gsim{\raise0.3ex\hbox{$>$\kern-0.75em\raise-1.1ex\hbox{$\sim$}}}

\def\beq{\begin{equation}}   
\def\eeq{\end{equation}}
\def\bea{\begin{eqnarray}}  
\def\eea{\end{eqnarray}}
\newcommand{\ba}{\begin{array}}
\newcommand{\ea}{\end{array}}
\def\nn{\nonumber}

\def\etmiss{\slashchar{E}_T}

\def\eps{\epsilon}
\def\tauptaum{\tau^+\tau^-}
\def\cnone{\widetilde \chi_1^0}

\def\tev{~{\rm TeV}}
\def\gev{~{\rm GeV}}

\def\what{\widehat}
\def\hpm{h^\pm}

\def\mhpm{m_{\hpm}}
\def\ie{{\it i.e.}}
\def\anti{\overline}
\def\br{BR}
\def\gam{\gamma}
\def\hsm{h_{SM}}

\def\mhh{m_{\hh}}
\def\mhl{m_{\hl}}
\def\hh{h_H}
\def\hl{h_L}
\def\hi{h_1}
\def\hii{h_2}

\def\ai{a_1}
\def\aii{a_2}
\def\mhi{m_{\hi}}
\def\mhii{m_{\hii}}

\def\mai{m_{\ai}}

\def\vev#1{\langle #1 \rangle}

\begin{document}

\begin{flushright}
UCD-2005-04 \\
LPT-Orsay-05-13
\end{flushright}
\vskip 12 truemm

\begin{center}

{\Large\bf Difficult Scenarios for NMSSM} \\
\vskip 2 truemm
{\Large\bf Higgs Discovery at the LHC}
\vskip 6 truemm

{\bf Ulrich Ellwanger}$^1$\footnote{Ulrich.Ellwanger@th.u-psud.fr}, 
{\bf John F. Gunion}$^2$\footnote{gunion@physics.ucdavis.edu},
{\bf Cyril Hugonie}$^1$\footnote{Cyril.Hugonie@th.u-psud.fr}
\vskip 6 truemm

$^1$Laboratoire de Physique Th\'eorique\\
Unit\'e Mixte de Recherche - CNRS - UMR 8627\\
Universit\'e de Paris XI, B\^atiment 210\\
F-91405 Orsay Cedex, France
\vskip 4 truemm

$^2$Department of Physics\\
University of California at Davis\\
Davis, CA 95616, U.S.A.

\end{center}
\vskip 8 truemm

\begin{abstract}
We identify scenarios not ruled out by LEP data in which NMSSM Higgs detection
at the LHC will be particularly challenging. We first review the `no-lose'
theorem for Higgs discovery at the LHC  that applies if Higgs bosons do not
decay to other Higgs bosons --- namely,  with $L=300\fbi$, there is always  one
or more `standard' Higgs detection channel with at least a $5\sigma$ signal.
However, we provide examples of no-Higgs-to-Higgs  cases for which all the
standard signals are no larger than $7\sigma$ implying that if the available $L$
is smaller or the simulations performed by ATLAS and CMS turn out to be overly
optimistic, all standard Higgs signals could fall below $5\sigma$ even in the
no-Higgs-to-Higgs part of NMSSM parameter space. In the vast bulk of NMSSM
parameter space, there will be Higgs-to-Higgs decays. We show that when such
decays are present it is possible for all the standard detection channels to
have very small significance. In most such cases,   the only strongly produced
Higgs boson is one with fairly SM-like couplings that decays to two lighter
Higgs bosons (either a pair of the lightest CP-even  Higgs bosons,  or, in the
largest part of parameter space, a pair of the lightest CP-odd Higgs bosons). A
number of representative bench-mark scenarios of this type are delineated in
detail and implications for Higgs discovery at various colliders are discussed.
\end{abstract}
\newpage

\section{Introduction}

One of the most attractive supersymmetric models is the Next to
Minimal Supersymmetric Standard Model (NMSSM)~\cite{allg} which
extends the MSSM by the introduction of just one singlet superfield,
$\what S$. When the scalar component of $\what S$ acquires a TeV scale
vacuum expectation value (a very natural result in the context of the
model), the superpotential term $\lam \what S \what H_u \what H_d$
generates an effective $\mu\what H_u \what H_d$ interaction for the
Higgs doublet superfields with $\mu=\lam \vev{S}$.  Such a term is
essential for acceptable phenomenology. No other SUSY model generates
this crucial component of the superpotential in as natural a fashion.
We also note that the LEP limits on Higgs bosons imply that the MSSM
must be very highly fine-tuned, whereas in the NMSSM parameter choices
consistent with LEP limits can be found that have very low
fine-tuning~\cite{Bastero-Gil:2000bw,Dermisek:2005ar}. Thus, the
phenomenological implications of the NMSSM at future accelerators
should be considered very seriously.

In the NMSSM,
the $h,H,A,\hpm$ Higgs sector of the MSSM is extended so that there
are three CP-even Higgs bosons ($h_{1,2,3}$,
$m_{h_1}<m_{h_2}<m_{h_3}$), two CP-odd Higgs bosons ($a_{1,2}$,
$m_{a_1}<m_{a_2}$) (we assume that CP is not violated in the Higgs
sector) and a charged Higgs pair ($h^\pm$). Hence, the Higgs
phenomenology in the NMSSM can differ significantly from the one in the
MSSM (see refs.~\cite{Ellwanger:2001iw,Miller:2003ay,Ellwanger:2003jt,
Ellwanger:2004gz,Miller:2004uh, Ellwanger:2004xm} for recent studies).

Our focus will be on NMSSM Higgs discovery at the LHC. 
An important question is
then the extent to which the no-lose theorem for MSSM Higgs boson
discovery at the LHC (see refs.~\cite{Denegri:2001pn,Schumacher:2004da} for
CMS and ATLAS plots, respectively) is retained when going to the
NMSSM; \ie\ is the LHC guaranteed to find at least one of the
$h_{1,2,3}$, $a_{1,2}$, $h^\pm$?

We will find that it is not currently possible to claim a no-lose
theorem for Higgs discovery in the NMSSM. This is due to the
importance of Higgs-to-Higgs decays in the NMSSM. Indeed, the no-lose
theorem for MSSM Higgs boson discovery at the LHC is based on Higgs
decay modes (hereafter referred to as `standard' modes) other than
Higgs-to-Higgs decays.\footnote{Higgs-to-Higgs decays do not create a
  problem for the CP-conserving MSSM no-lose theorem due to the
  constrained nature of the MSSM Higgs sector. Relations among the
  MSSM Higgs boson masses are such that Higgs pair decays are only
  possible if $m_A$ is quite small. In this part of parameter space,
  the $H$ is SM-like and $H\to AA$ decays can be dominant. However,
  when $m_A$ is small, the $h$ also has small mass {\it and} the $Z\to
  hA$ coupling is large.  As a result, $Z\to hA$ pair production would
  have been detected at LEP.} The importance of such decays was first
noted in~\cite{Gunion:1996fb} and later pursued
in~\cite{Dobrescu:2000yn,Dobrescu:2000jt}.  Correspondingly, the
parameter space of the NMSSM can be decomposed into the following
three regions:

a) An (actually fairly small) region where, for kinematical reasons,
Higgs-to-Higgs decays are forbidden. Here, Higgs detection in the
NMSSM proceeds via the standard discovery modes, with possibly reduced
couplings and altered branching ratios
with respect to the MSSM. In a first exploration of this
part of the NMSSM parameter space \cite{Gunion:1996fb}, significant
regions were found such that the LHC would not detect any of the NMSSM
Higgs bosons. Since then, however, there have been improvements in
many of the detection modes (and the addition of new ones). As a result
\cite{Ellwanger:2001iw}, if the neutral NMSSM
Higgs bosons do not decay to other Higgs bosons, then the LHC is guaranteed
to discover at least one of them for an integrated luminosity of
$L=300\fbi$ at both the ATLAS and the CMS detectors.

b) The largest region of the NMSSM parameter space is the part where
Higgs-to-Higgs decays are kinematically possible, but where the standard
discovery modes are still sufficient for the detection of at least one Higgs
boson at the LHC.

c) For a small part of the NMSSM parameter space Higgs-to-Higgs decays
are dominant for the Higgs bosons with substantial production cross
sections, and the standard discovery modes do not yield a $5\sigma$
signal (even for integrated luminosity of $L=300\fbi$ {\it and} after
combining modes) for any of the Higgs bosons. In
Refs.~\cite{Ellwanger:2003jt,Ellwanger:2004gz}, we presented a
selection of benchmark points with these characteristics. However,
since then the expected statistical significances at the LHC in the
standard discovery channels have improved and some of these points
would now give a $5\sigma$ signal.  (On the other hand,
in~\cite{Ellwanger:2003jt,Ellwanger:2004gz} we were somewhat
optimistic with respect to LEP limits on Higgs bosons with masses
below $115\gev$ and unconventional decay modes.)

In this paper we repeat these studies, updating the LEP constraints
and the expected statistical significances for the standard discovery
modes at the LHC. Once again, we find a region in the NMSSM parameter
space of type c) above.  In section~\ref{higgstohiggs}, we present
new benchmark points for which the primary decaying neutral Higgs
boson ($\hh$) has strong coupling to gauge bosons and has mass in the
range $[90\gev,150\gev]$ but decays almost entirely to a pair of even
lighter secondary Higgs states ($\hl\hl$).  Both the primary and
secondary Higgs bosons will have escaped LEP searches and will be
impossible to observe at the LHC in the standard modes.  The benchmark
points presented are chosen to represent a range of $\mhh$ and $\mhl$
possibilities and a variety of possible $\hl$ decays.

The outline of the paper is as follows: In section~\ref{scanning}, in
preparation for our discussions, we define the NMSSM model and its
parameters. There, we also review the program NMHDECAY
~\cite{Ellwanger:2004xm} employed for this study, and specify our
precise scanning procedures for the Higgs discovery studies.

In section~\ref{nohiggstohiggs}, we review the conclusions of
\cite{Ellwanger:2001iw} regarding the above region a) of the NMSSM
parameter space. These remain unchanged: assuming an integrated
luminosity of $L=300\fbi$, one can establish a no-lose theorem for the
very restricted part of parameter space where there are no decays of
neutral Higgs bosons to other Higgs bosons. The statistical
significances as a function of the charged Higgs mass, and the
properties of two relatively difficult points in this region of
parameter space (but still with a $5\sigma$ signal in at least one of
the standard discovery modes) are presented.

In section~\ref{higgstohiggs}, we discuss general properties of points
for which one or more Higgs-to-Higgs decays are allowed and, as a
result, discovery of a Higgs boson in one of the standard modes is not
possible. We present eight new benchmark points, discuss their properties,
and show how Higgs-to-Higgs decays can lead to very small signals in
all the usual LHC Higgs discovery channels for these points.

In section~\ref{colliders}, we will discuss the nature and detectability of the
collider signals for the Higgs pair decay modes, especially focusing on the
difficulties at hadron colliders such as the Tevatron and LHC. Notably we
propose that the LHC may be able to detect Higgs-pair final states using the
$WW\to \hh \to \hl\hl \to jj\tauptaum$ production/decay mode, and discuss its
properties and possible cuts. Final conclusions are
given in section~\ref{conclusions}.

\section{The model, the NMHDECAY program and the scanning procedures}
\label{scanning}

We consider the simplest version of the NMSSM
\cite{allg}, where the term $\mu \widehat
H_1 \widehat H_2$ in the superpotential of the
MSSM is replaced by (we use the notation $\widehat
A$ for the superfield and $A$ for its scalar
component field)
\begin{equation}\label{2.1r}
\lambda \widehat H_1 \widehat H_2 \widehat S\ + \ \frac{\kappa}{3} \widehat S^3
\ \ ,
\end{equation}
\noindent so that the superpotential is scale invariant. 
The associated trilinear soft terms are
\beq \label{1.2} \lambda A_{\lambda} S H_u H_d + \frac{\kappa}
{3} A_\kappa S^3\,. \eeq
The final two Higgs-sector input parameters are
\beq \label{1.3} \tan \beta =\ \left< H_u \right>/ \left< H_d \right>\ ,
 \ \mu_\mathrm{eff} = \lambda \left< S \right>\ . \eeq
These, along with $\mz$, can be viewed as determining the
three \susy\ breaking masses squared for $H_u$, $H_d$
and $S$ appearing in the soft-\susy-breaking terms
\begin{equation}\label{2.2r}
m_{H_1}^2 H_1^2\ +\ m_{H_2}^2 H_2^2\ +\ m_S^2 S^2\ 
\end{equation}
\noindent through the three minimization equations of the
scalar potential.
Thus, we make no assumption of ``universal'' soft terms.

In short, as compared two independent parameters in the Higgs
sector of the MSSM (often chosen as $\tan \beta$ and $M_A$), 
the Higgs sector of the NMSSM is described by
the six parameters
\beq \label{6param}
\lambda\ , \ \kappa\ , \ A_{\lambda} \ , \ A_{\kappa}, \ \tan \beta\ ,
\ \mu_\mathrm{eff}\ .
  \eeq
  We will choose sign conventions for the fields such that $\lambda$
  and $\tan\beta$ are positive, while $\kappa$, $A_\lambda$,
  $A_{\kappa}$ and $\mu_{\mathrm{eff}}$ should be allowed to have
  either sign.  We will perform a scan over these parameters using the
  publicly available program NMHDECAY \cite{Ellwanger:2004xm}.  For
  any choice of the above parameters and other soft-\susy-breaking
  parameters that affect radiative corrections and Higgs decays, 
NMHDECAY performs the following tasks:
\ben
\item It computes the masses and couplings of all the physical Higgs
  and sparticle states.  We only retain points for which all Higgs and
  squark/slepton masses-squared are positive.

\item It checks whether the running Yukawa couplings encounter a Landau
singularity below the GUT scale. 
In our scans, we eliminate such cases.

\item NMHDECAY checks whether the physical minimum (with all vevs 
non-zero) of the scalar potential is deeper than the local unphysical minima
with vanishing $\vev{H_u}$, $\vev{H_d}$ or $\vev{S}$. 
We keep only parameter choices for which
the minimum with all vevs non-zero is the true minimum.

\item It computes the branching ratios into two particle final states
(including char\-ginos and neutralinos) of all Higgs particles.
Currently, squark
and slepton decays of the Higgs are not computed. 

\item It checks whether the Higgs masses and couplings violate any bounds
from negative Higgs searches at LEP, including many quite
unconventional channels that are relevant for the NMSSM Higgs sector.
It also checks the bound on the invisible $Z$ width (possibly violated
for light neutralinos). In addition, NMHDECAY checks
the bounds on the lightest chargino and on neutralino pair
production. 
Parameter choices that conflict with LEP bounds are eliminated
in the scans discussed below.

\een
Points that pass the requirements of items 1 through 3 above 
define the set of ``physically acceptable''
parameter choices.
Our scans will be for randomly chosen
parameter values in the following ranges:
\bea
&10^{-4}\leq\lam\leq 0.75;\quad
-0.65\leq\kappa\leq 0.65;\quad
1.6\leq\tanb\leq 54& \nonumber\\
&-1\tev\leq \mueff,A_\lam,A_\kappa\leq +1\tev\,.&
\label{ranges}
\eea

In the gaugino sector, we chose $M_2=1\tev$ (at low scales). We assume
universal gaugino masses at the coupling
constant unification scale,
leading to $M_1\sim 500\gev$ and $M_3\sim 3\tev$.
Thus, the lightest neutralino can only be significantly
lighter than $500\gev$ if it is mainly singlino
or (when $\mueff$ is relatively small) higgsino.
For the chosen $M_{1,2,3}$ values, LHC detection of the gauginos
will be quite difficult and decay 
of Higgs bosons to gauginos, including the invisible $\cnone\cnone$
channel, will in most cases be negligible.

Current lower limits from LEP and the Tevatron imply that squarks and
sleptons must be at least moderately heavy.  As a result, the Higgs
bosons with substantial $WW/ZZ$ coupling (which are predicted to have
masses below about $150\gev$) cannot decay to squarks and sleptons.
We will choose squark/slepton parameters in the TeV range. In this
case decays to squarks and sleptons are unimportant or absent also for
the heavier Higgs states.  Specifically, we choose
$m_Q = m_U=m_D=m_L=m_E \equiv \msusy= 1$ TeV for the soft-\susy-breaking
masses for all generations.  This means that squarks and sleptons will
be at the edge of the LHC discovery reach and that Higgs boson
detection might be the only new physics signal.\footnote{In addition, if
  the sparticles have large masses their contributions to the loop
  diagrams inducing Higgs boson production by gluon fusion and Higgs
  boson decay into $\gamma \gamma$ are negligible. }

For the trilinear soft-\susy-breaking squark parameters, we choose
$A_U=A_D=1.5\tev$ for all generations, including the third
generation.  We recall that the light Higgs mass is maximized by
choosing the parameter $A_T$ so that
\begin{equation}\label{2.4r}
X_T \equiv \frac{A_T-\mueff\cot\beta}{\sqrt{\msusy^2+\mt^2}}
\end{equation} 
\noindent  
%
takes the value $X_T = \sqrt{6}$ (so called maximal mixing). 
For $A_T=1.5\tev$, $\msusy=1\tev$
and $\mt=175\gev$, one finds $X_T\sim 1.5$, which yields lower Higgs
masses than maximal mixing and so we are not typically choosing the
most difficult scenarios that we could.

Overall, for the squark and gaugino masses above,
it might be that direct detection of the supersymmetric particles
would not be possible and that the only new signal would be
the detection of a Higgs boson. This makes the issue
of whether or not at least one of the NMSSM Higgs bosons is
guaranteed to be detectable at the LHC of vital importance.

Finally, in our scan over parameter space, 
we restrict ourselves to the region $\mhpm > 155\gev$. 
For moderate $\tanb$, this means  
that $t \to h^\pm b$ and other possible charged Higgs signals
would not reach the $\geq 5\sigma$ level  at the LHC (see later
discussion).

\section{LHC prospects when Higgs-to-Higgs decays are forbidden}
\label{nohiggstohiggs}

In this section, we update the no-lose theorem
for NMSSM Higgs detection for the small portion of parameter
space in which there are no decays of neutral Higgs bosons
to other Higgs bosons.
In the absence of such decays, the most relevant modes for 
detecting the neutral NMSSM
Higgs bosons are those that have been earlier considered for the SM
and for the MSSM~\cite{Kinnunen:1997,unknown:1999fr,Zeppenfeld:2000td,Zeppenfeld:2002ng}.  
These are (with $\ell=e,\mu$)

1) $g g \to h/a \to \gamma \gamma$;\par
2) associated $W h/a$ or $t \bar{t} h/a$ production with 
$\gamma \gamma\ell^{\pm}$ in the final state;\par
3) associated $t \bar{t} h/a$ production with $h/a \to b \bar{b}$;\par
4) associated $b \bar{b} h/a$ production with $h/a \to \tau^+\tau^-$;\par
5) $g g \to h \to Z Z^{(*)} \to$ 4 leptons;\par
6) $g g \to h \to W W^{(*)} \to \ell^+ \ell^- \nu \bar{\nu}$;\par
7) $W W \to h \to \tau^+ \tau^-$;\par
8) $W W \to h\to W W^{(*)}$.\par
\noindent In addition to these, we also include in our work the mode \par
9) $WW\to h \to invisible$. \par
\noindent LHC sensitivity to this mode has been studied by the
ATLAS \cite{unknown:1999fr,Cavalli:2002vs} and CMS
\cite{Abdullin:2003,nikmaz:2002,Cavalli:2002vs}
collaborations. In the present
study, we employed the results of ref.~\cite{Abdullin:2003} (fig.~25) 
that covers the Higgs
mass range  $[100\gev,400\gev]$ and
includes systematic uncertainties. Details regarding how we treat
all these modes are given in the Appendix. 

\begin{figure}[h]
\begin{center}
\epsfig{file=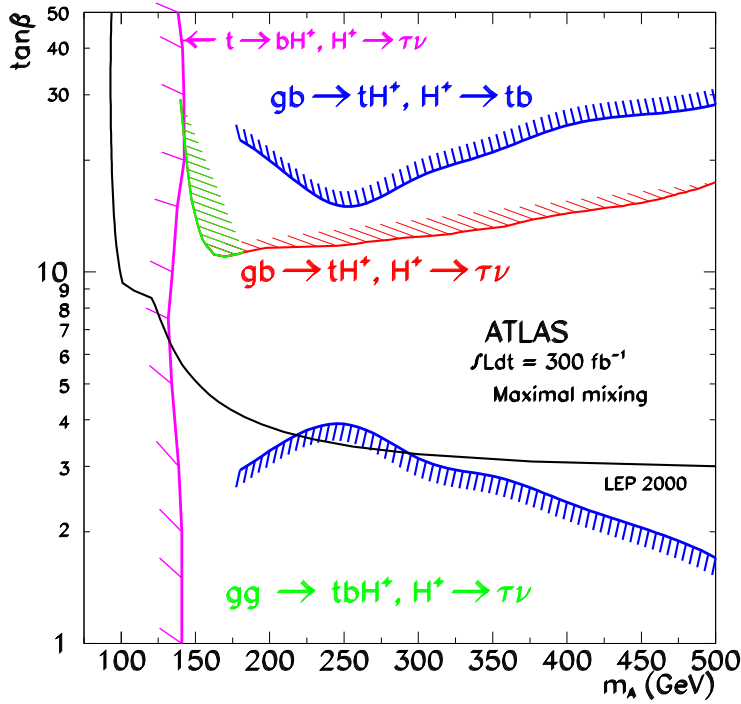,width=10.5cm}
\caption{We display the contours for $5\sigma$ charged Higgs
detection from~\cite{Assamagan:2004gv}.
}
\label{newatlas}
\end{center}
\end{figure}

As regards the charged Higgs boson, it is well established
(for early studies, see refs.~\cite{unknown:1999fr,Kinnunen:1997}) 
that if $t\to \hpm b$ decays are kinematically allowed
then the $\hpm$ will be relatively easily 
discovered in the decay
of the top quark in $t\anti t$ events. 
These earlier studies have been updated by ATLAS
in~\cite{Assamagan:2004gv}.
The resulting $L=300\fbi$ plot appears in fig.~\ref{newatlas}.
The least sensitivity to a charged Higgs boson
occurs for $\tanb\sim 4\div 10$
where $5\sigma$ is only attained for the MSSM
parameter $m_A\lsim 135\gev$,
corresponding to $\mhpm\lsim 155\gev$.
Once $\mhpm\gsim 210\gev$, the Higgs-to-Higgs decay
mode $\hpm\to\wpm h_1$ 
is typically kinematically allowed and in the NMSSM 
can have substantial
branching ratio. Thus, in the NMSSM context the limits of
fig.~\ref{newatlas} on
the $\hpm$ in the region $\mhpm\gsim 210\gev$ do not apply.

We will study the complementarity between charged Higgs detection and
neutral Higgs detection in the standard modes 1) -- 9).  We find
that the smaller the lower limit on $\mhpm$ for which we assume good
significance for $t\to \hpm b$ detection, the smaller can be the
minimum statistical significance for the neutral Higgs detection
modes. 

In Ref.~\cite{Ellwanger:2001iw}, a
partial no-lose theorem for NMSSM Higgs boson discovery at the LHC (when
Higgs-to-Higgs decays are forbidden) was
established based on modes 1) - 8) above. There, we estimated the statistical
significances ($N_{SD}=S/\sqrt B$) for modes 1) - 8).  For these results,
it was especially critical that the $t\anti t h$ with $h\to b\anti b$ and
$WW$-fusion modes [3) and 7), respectively] were included.
Also important was mode 4), $b\anti b h$ with $h\to \tauptaum$.
In the case of $t\anti t h$ with $h\to b\anti b$, we used the 
experimental study done by V. Drollinger at our request (see the
Appendix) that extends results for this mode to Higgs masses as large
as $150\gev$.  The conclusion of ref.~\cite{Ellwanger:2001iw} was
that, for an integrated luminosity of $L=300\fbi$ at the LHC, all the
surviving points yielded $N_{SD}>10$ after combining all modes.  This
means that NMSSM Higgs boson discovery by just one detector with
$L=300~{\rm fb}^{-1}$ is essentially guaranteed for those portions of
parameter space for which Higgs boson decays to other Higgs bosons 
 are kinematically forbidden.

For the present paper, we have repeated the scan described above with
the latest available LEP constraints (as incorporated in NMHDECAY
\cite{Ellwanger:2004xm} --
see references therein) and with the latest ATLAS and CMS results
for the discovery channels 1) -- 9) (see the Appendix)
and most recent $5\sigma$ curve for the $\hpm$ as given
in fig.~\ref{newatlas}.  For each Higgs state, 
we calculated all branching ratios using NMHDECAY.
 We then estimated the expected statistical significances
at the LHC in all Higgs boson detection modes 1)
-- 9) by rescaling results for the SM Higgs boson
and/or the MSSM $h, H$ and/or $A$. The
rescaling factors for the CP-even $h_i$ are determined by $R_i$, $t_i$,
$b_i=\tau_i$, $g_i$ and $\gam_i$, the ratios of the $VVh_i$,
$t\anti t h_i$, $b\anti b h_i$ (or $\tau^+\tau^- h_i$), $ggh_i$ 
and $\gam\gam h_i$ couplings, respectively, to those of a SM Higgs
boson of the same mass.  
Of course $|R_i| < 1$, but $t_i$, $b_i$, $g_i$ and $\gam_i$ 
can be larger, smaller or even differ in sign with
respect to the SM. The reduced couplings for the CP-odd Higgs bosons
(denoted by primes) are as follows:
$R_i'=0$ at tree-level; $t'_j$ and $b'_j$ are the
ratios of the $i\gamma_5$ couplings for $t\bar{t}$
and $b\bar{b}$, respectively, relative to SM-like
strength.  The quantities $g_i'$ and $\gam_i'$
are the ratios of the $\eps\times\eps'$ ($\eps$ and $\eps'$ being
the polarizations of the gluons or photons) $gg a_i$ or $\gam\gam a_i$
coupling strength to the $\eps\cdot \eps'$ $gg \hsm$ or $\gam\gam\hsm$
coupling strength for $m_{a_i}=m_{\hsm}$.
A detailed discussion of the procedures
for rescaling SM and MSSM simulation results for
the statistical significances in channels 1) -- 9)
is given in the Appendix. We will now summarize
the results of this new no-Higgs-to-Higgs scan.

Only a few parameter choices are such that decays of a neutral
Higgs boson to any other Higgs boson
are all forbidden when $\mhpm\geq 155\gev$ is required.
After restricting to such parameters,
it is extraordinarily difficult to locate points 
that do not yield large statistical significance for LHC discovery
of at least one Higgs boson while at the same time
LEP constraints are not violated. 
We obtained a sample of $2455$ points
that had LHC significance (for $L=300\fbi$)
in channels 1) -- 9) below $10\sigma$. Most of these points
had $4\lsim \tanb\lsim 10$ --- high $\tanb$ enhances
production cross sections for some of the Higgs bosons
and will typically lead to visible signals. All points had
LHC statistical significance above $5\sigma$, thus establishing
a no-lose theorem for points chosen consistent with LEP constraints
and absence of Higgs-to-Higgs decays. Statistics on the
important channels for these $2455$ points are summarized in
table~\ref{nohtoh}.
Note the importance of the channels 3), 4) and 7)
for these most difficult cases.

\begin{table}[h]
\begin{center}
\begin{tabular}{|l|r|r|r|r|r|r|r|r|r|}
\hline
Channel with highest $S/\sqrt B$ & 1 & 2 & 3 &4 & 5 & 6 & 7 & 8 & 9 \\
\hline
No. of points & 0 & 0 & 343 & 132 & 0 & 1 & 1979 & 0 & 0\\
\hline
\end{tabular}
\end{center}
\caption{\label{nohtoh} Most important channel for detecting the $2455$ 
no-Higgs-to-Higgs-decays points that were most difficult for LHC detection.}
\end{table}

\begin{figure}[ht]
\begin{center}
\epsfig{file=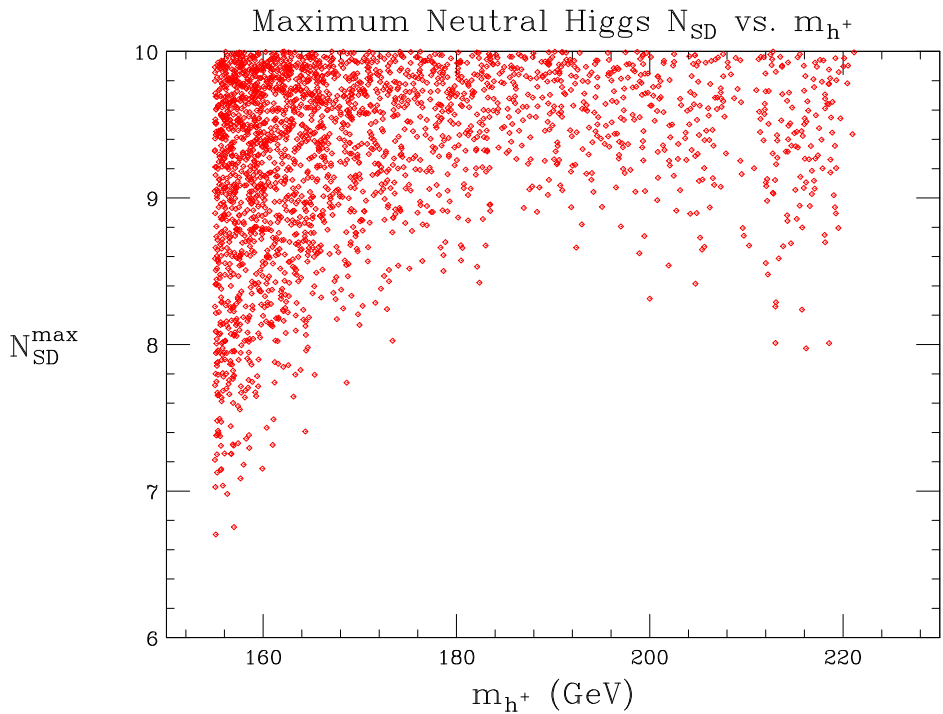,angle=90,width=12.5cm}
\caption{$N_{SD}^{\rm max}$ for the neutral Higgs
  bosons vs. $\mhpm$ for $\mhpm\geq 155\gev$. 
Here, $N_{SD}^{\rm max}$ is the largest of
the significances for any one of the neutral Higgs bosons in channels
1) - 9). We assume LHC luminosity of $L=300\fbi$.
}
\label{sigvsmhpm}
\end{center}
\end{figure}

In fig.~\ref{sigvsmhpm}, we give a scatter plot
of the largest statistical significance
achieved for a single neutral Higgs boson, $N_{SD}^{\rm max}$, as a function
of $\mhpm$ for the above $2455$ points. 
(The absence of points with $\mhpm\gsim 210\gev$ is 
due to the fact that above this scale Higgs-to-Higgs decays,
typically $\aii \to Z \hi$ and $\hpm\to \wpm h_1$, would be allowed.) 
We see that the larger
the value of $\mhpm$  for which a $5\sigma$ signal can be
established using $t\anti t$ production with $t\to \hpm b$ decay,
the larger the minimum possible value of the neutral
Higgs bosons' $N_{SD}^{\rm max}$. This means that the ATLAS and CMS
groups should work to maximize sensitivity to charged Higgs production
as well as to neutral Higgs production. 

The point yielding the very lowest
LHC statistical significance had the following parameters,
\bea &\lam=0.0163;\quad \kappa=-0.0034;\quad \tanb=5.7;&\nonumber\\
&\mueff=-284\gev;\quad A_\lam=-70\gev;\quad A_\kappa=-54\gev\,, &
 \eea
 which yielded $\mhpm\sim 155\gev$ and neutral Higgs boson properties as given
in table~\ref{higgsprop}.  Other points among the $2455$ are similar in that
the Higgs masses are closely spaced and below or at least not far above the
$WW/ZZ$ decay thresholds, the CP-even Higgs bosons tend to share the $WW/ZZ$
coupling strength (indicated by $R_i$ in the table), couplings to $b\anti b$ of
all Higgs bosons (the $b_i$ or $b_i'$ in the table) are not very enhanced, and
couplings to $gg$ and $\gam\gam$ (the $g_i$ or $g_i'$ and $\gam_i$ or $\gam_i'$
in the table) are suppressed relative to the SM Higgs strength.  The most
visible processes for this point had $N_{SD}$ values above $6$. These  were the
$WW\to h_2\to\tauptaum$, $WW\to h_3\to \tauptaum$ and $t\anti t h_2\to t\anti t
b\anti b$ channels.  Overall, we have a quite robust LHC no-lose theorem for
NMSSM parameters such that LEP constraints are passed and Higgs-to-Higgs decays
are not allowed.

\begin{table}[ht]

 \begin{center}
 \footnotesize
 \begin{tabular}{|l|r|r|r|r|r|}
 \hline
 Higgs           & $  h_1$ & $  h_2$ & $  h_3$ & $  a_1$ & $  a_2$ \\
 \hline
 Mass (GeV)      & $  99$ & $ 114$ & $ 145$ & $  98$ & $ 134$ \\ 
 \hline
 $R_i$           & $ 0.49$ & $ 0.72$ & $-0.48$ & $    -$ & $    -$ \\
 \hline
 $t_i$ or $t_i'$ & $ 0.46$ & $ 0.65$ & $-0.64$ & $-0.01$ & $ 0.18$ \\
 \hline
 $b_i$ or $b_i'$ & $ 1.71$ & $ 3.23$ & $ 4.49$ & $ 0.36$ & $ 5.59$ \\
 \hline
 $g_i$ or $g_i'$ & $ 0.41$ & $ 0.56$ & $ 0.79$ & $ 0.02$ & $ 0.14$ \\
 \hline
 $\gam_i$ or $\gam_i'$
                 & $ 0.51$ & $ 0.75$ & $ 0.43$ & $ 0.01$ & $ 0.10$ \\
 \hline
 $\br(h_i~or~a_i\to b\anti b)$
                 & $ 0.91$ & $ 0.90$ & $ 0.88$ & $ 0.92$ & $ 0.91$ \\
 \hline
 $\br(h_i~or~a_i\to \tauptaum)$
                 & $ 0.08$ & $ 0.08$ & $ 0.09$ & $ 0.08$ & $ 0.09$ \\
 \hline
 Chan. 1) $S/\sqrt B$
                 &  0.00   &  0.22   &  0.20   &  0.00   &  0.00   \\
 \hline
 Chan. 2) $S/\sqrt B$
                 &  0.42   &  0.80   &  0.15   &  0.42   &  0.00   \\
 \hline
 Chan. 3) $S/\sqrt B$
                 &  3.52   &  6.25   &  5.39   &  3.52   &  5.39   \\
 \hline
 Chan. 4) $S/\sqrt B$
                 &  0.73   &  1.26   &  3.86   &  1.26   &  3.86   \\
 \hline
 Chan. 5) $S/\sqrt B$
                 &  0.00   &  0.15   &  1.00   &  $-$    &  $-$    \\
 \hline
 Chan. 6) $S/\sqrt B$
                 &  0.00   &  0.00   &  0.80   &  $-$    &  $-$    \\
 \hline
 Chan. 7) $S/\sqrt B$
                 &  0.00   &  6.70   &  6.54   &  $-$    &  $-$    \\
 \hline
 Chan. 8) $S/\sqrt B$
                 &  0.00   &  0.20   &  0.25   &  $-$    &  $-$    \\
 \hline
 All-channel $S/\sqrt B$
                 & 3.61    &  9.29   &  9.41   &  3.76   &  6.63   \\
 \hline
 \end{tabular}
 \end{center}

 \caption{\label{higgsprop} Properties of the neutral NMSSM Higgs bosons for the
 most difficult no-Higgs-to-Higgs-decays LHC point. In the table, $R_i=g_{h_i
 VV}/g_{h_{SM} VV}$,  $t_i=g_{h_i t\anti t}/g_{h_{SM} t\anti t}$,
 $b_i=g_{h_ib\anti b}/g_{h_{SM} b\anti b}$, $g_i=g_{h_i gg}/g_{h_{SM} gg}$ and
 $\gam_i=g_{h_i \gam\gam}/g_{h_{SM} \gam\gam}$   for $m_{h_{SM}}=m_{h_i}$.
 Similarly, $t_i'$ and $b_i'$ are the $i\gam_5$ couplings of $a_i$  to $t\anti
 t$ and $b\anti b$ normalized relative to the scalar  $t\anti t$ and $b\anti b$
 SM Higgs couplings and $g_i'$ and $\gam_i'$ are the  $a_i gg$ and $a_i\gam\gam$
 $\eps\times\eps'$ couplings relative to the $\eps\cdot\eps'$ coupling of the SM
 Higgs.}

\end{table}

\begin{table}[h!]

 \begin{center}
 \footnotesize
 \begin{tabular}{|l|r|r|r|r|r|}
 \hline
 Higgs           & $  h_1$ & $  h_2$ & $  h_3$ & $  a_1$ & $  a_2$ \\
 \hline
 Mass (GeV)      & $ 113.$ & $ 126.$ & $ 203.$ & $ 150.$ & $ 202.$ \\ 
 \hline
 $R_i$           & $ 0.75$ & $ 0.66$ & $ 0.01$ & $    -$ & $    -$ \\
 \hline
 $t_i$ or $t_i'$ & $ 0.74$ & $ 0.65$ & $ 0.22$ & $-0.01$ & $ 0.12$ \\
 \hline
 $b_i$ or $b_i'$ & $ 1.38$ & $ 1.18$ & $-8.11$ & $-0.78$ & $ 8.22$ \\
 \hline
 $g_i$ or $g_i'$ & $ 0.72$ & $ 0.63$ & $ 0.37$ & $ 0.01$ & $ 0.08$ \\
 \hline
 $\gam_i$ or $\gam_i'$
                 & $ 0.76$ & $ 0.65$ & $ 0.93$ & $ 0.06$ & $ 0.05$ \\
 \hline
 $\br(h_i~or~a_i\to b\anti b)$
                 & $ 0.87$ & $ 0.80$ & $ 0.86$ & $ 0.91$ & $ 0.90$ \\
 \hline
 $\br(h_i~or~a_i\to \tauptaum)$
                 & $ 0.08$ & $ 0.08$ & $ 0.09$ & $ 0.09$ & $ 0.10$ \\
 \hline
 Chan. 1) $S/\sqrt B$
                 &  1.91   &  2.05   &  0.00   &  0.00   &  0.00   \\
 \hline
 Chan. 2) $S/\sqrt B$
                 &  4.99   &  4.02   &  0.00   &  0.00   &  0.00   \\
 \hline
 Chan. 3) $S/\sqrt B$
                 &  7.84   &  5.95   &  0.00   &  0.00   &  0.00   \\
 \hline
 Chan. 4) $S/\sqrt B$
                 &  0.20   &  0.20   &  7.40   &  0.05   &  7.40   \\
 \hline
 Chan. 5) $S/\sqrt B$
                 &  1.33   &  2.84   &  0.31   &  $-$    &  $-$    \\
 \hline
 Chan. 6) $S/\sqrt B$
                 &  0.00   &  2.23   &  0.19   &  $-$    &  $-$    \\
 \hline
 Chan. 7) $S/\sqrt B$
                 &  6.67   &  7.62   &  0.00   &  $-$    &  $-$    \\
 \hline
 Chan. 8) $S/\sqrt B$
                 &  1.14   &  3.85   &  0.00   &  $-$    &  $-$    \\
 \hline
 All-channel $S/\sqrt B$
                 & 11.73   &  11.91  &  7.41   &  0.05   &  7.40   \\
 \hline
 \end{tabular}
 \end{center}

 \caption{\label{higgsprop2} Properties of the neutral NMSSM Higgs bosons
 for the second interesting point described in the text. Notation as in
 table~\ref{higgsprop}.}

\end{table}

Another example point of possible interest is that giving
the weakest signals when the charged Higgs mass is near the upper
end of the spectrum for which the $\hpm$ does not decay to other Higgs
bosons. (This requirement is what restricts the upper range of $\mhpm$
in fig.~\ref{sigvsmhpm}.)
The parameters for this point are:
\bea &\lam=0.0379;\quad \kappa=-0.0238;\quad \tanb=8.25;&\nonumber\\
&\mueff=-119\gev;\quad A_\lam=-116\gev;\quad A_\kappa=-102\gev\,, &
 \eea
yielding a charged Higgs mass of $216\gev$ and neutral Higgs
properties and statistical significances as listed in
table~\ref{higgsprop2}. Note the strong signals in the
$WW\to h_{1,2} \to \tauptaum$, $t\anti t h_{1,2} \to t\anti t b\anti b$ and 
$b\anti b h_3,a_2 \to b\anti b\tauptaum$ channels.

These particular points in parameter space illustrate the general
conclusion that it will be important that all the neutral Higgs detection modes
that have been simulated by ATLAS and CMS really achieve their full
$L=300\fbi$ potential.  If the effective luminosity accumulated in
modes 3), 4) and 7) were to all 
fall below $L\sim 100\fbi$, then all single
channel statistical significances for the most marginal points (as
exemplified by the two tabulated points) would fall below $5\sigma$.
Channel combination would be required to reach the $\geq 5\sigma$
level.

\section{LHC prospects when Higgs-to-Higgs decays are allowed}
\label{higgstohiggs}

In this section we will consider
the part of the parameter space complementary to that
scanned in section~\ref{nohiggstohiggs}.
To be precise, we require that {\it at least one} of the
following decay modes be kinematically allowed for some $h$ and or $a$:
\begin{eqnarray}
& i) \ h \to h' h' \; , \quad ii) \ h \to a a \; , \quad iii) \ h \to h^\pm
h^\mp \; , \quad iv) \ h \to a Z \; , \nonumber \\
& v) \ h \to h^\pm W^\mp \; , \quad vi) \ a' \to h a \; , \quad vii) \ a \to h
Z \; , \quad viii) \ a \to h^\pm W^\mp \; .
\end{eqnarray}
(Recall that we do not consider $\hpm\to \wpm h$ or $\hpm\to \wpm a$ 
in defining the Higgs-to-Higgs decay parameter region.)
The branching ratios for all these decays are computed by NMHDECAY.
As in the previous section, we also allow for (but do not require)
Higgs decays to gauginos. The large gaugino masses
we employ imply that such decays are never important for the scans
discussed in this paper.

For most of these points it turns out that $5\sigma$ discovery of a neutral 
Higgs boson in at least one of the modes 1) -- 9) is still possible.
The number of parameter space points
for which one or more of the decays $ i) - viii)$
is allowed, but $5\sigma$ discovery of a neutral Higgs boson in
modes 1) -- 9) is not possible, represents less than $1\%$ of the physically
acceptable points; in our scan we have found $3480$
such points. In one sense, this small percentage is encouraging in that it
implies that the standard LHC detection modes 1) -- 9) suffice for most
of randomly chosen parameter points. However, it should be
noted that the fraction of points for which modes 1) -- 9)
suffice will decrease rapidly as the assumed LHC 
integrated luminosity is reduced.  Further, the difficult
parameter points are preferred on the basis of keeping fine-tuning
modest in size~\cite{Dermisek:2005ar}. (Modest fine-tuning
means that $\mz$ is not very sensitive
to GUT scale choices for the soft-\susy-breaking parameters.)  

The parameters associated with these points 
for which all NMSSM Higgs bosons escape LEP detection
and LHC detection in modes 1) -- 9) occur throughout the broad
range defined in eq.~(\ref{ranges}).
The scenarios associated with these
points have some generic properties of considerable interest 
that make them worthy of further study.
First, for all these $3480$ points, the $h_3$ and $a_2$
are so heavy that they  will only be detectable if a super high
energy LC is eventually built so that $e^+e^-\to
Z\to h_3 a_2$ is possible, implying that LHC Higgs detection must rely on
the lighter $\hi$, $\hii$ and $\ai$ states. 
The NMSSM parameter
choices for which the latter cannot be detected at the LHC
in the standard modes are such that there is a light, fairly SM-like CP-even
Higgs boson ($h_1$ or $h_2$) that
decays mainly to two lighter CP-odd or CP-even Higgs bosons ($h_{1,2}\to
a_1a_1$ or $h_2\to h_1h_1$). We will denote the parent SM-like
CP-even Higgs boson by $\hh$ and the daughter Higgs boson that appears
in the pair decay by $\hl$.

We should discuss how it is that
a light $\hl$ will have escaped LEP detection.
Consider the case of $\hl=a_1$.
First, sum rules require that the $Zh_1 a_1$ ($Zh_2 a_1$) coupling
is small when the $h_1$ ($h_2$) coupling is
near SM strength, implying that discovery in the $\epem\to Z^*\to h_1a_1$
($\epem\to Z^*\to h_2a_1$) mode will not be possible.
Second, $e^+e^-\to Z^*\to h_2a_1$ ($h_1a_1$) LEP 
constraints can be evaded in the NMSSM since the $a_1$ can
have sufficient singlet component that
the $Zh_2 a_1$ ($Zh_1 a_1$) coupling is small when
the $h_1$ ($h_2$) is SM-like. 
For scenarios
in which the $h_2$ is SM-like and decays
primarily via $h_2\to h_1 h_1$, the $h_1$
is not observed at LEP because of its weak $ZZ$
coupling, while the $h_2$ mass is beyond the reach of LEP.
\footnote{A similar situation arises 
in the case of a CP-violating MSSM Higgs
sector~\cite{Carena:2002bb}. 
There, the three Higgs bosons are mixed and parameter choices
for which $h_2\to h_1h_1$ decays are dominant can be found
for which LEP constraints would not apply despite the fact that
the $h_1$ is quite light.}  

As we review the properties of specific bench mark points, it is
useful to keep in mind the fact that detection of Higgs-pair final
states at the LHC might be possible in certain cases.  In particular,
in ~\cite{Ellwanger:2003jt,Ellwanger:2004gz} we proposed that the LHC
may be able to detect Higgs-pair final states using the $WW\to \hh\to
\hl\hl\to jj\tauptaum$ production/decay mode.  By this we mean that
one of the final $\hl$ Higgs bosons is required to decay to a
$\tauptaum$ pair where we identify the $\tau$'s through their leptonic
decays to electrons and muons. The other final $\hl$ is required to
decay to either $b\anti b$ (which we identify as jets) or, if $b\anti
b$ is not a kinematically allowed decay, to $jj$ or (with much smaller
identification efficiency) $\tauptaum$ where the $\tau$'s of this
second pair would be tagged via their decays to single jets plus
neutrinos.  However, for a small fraction of the $3480$ points, $\hh
\to a_1a_1$ decays are prominent but $m_{a_1}\leq 2\mtau$.  For
another small fraction, the $\hl$ has suppressed couplings to
$b\anti b$ and $\tauptaum$. In either case, $\tau$ triggering does not
work and NMSSM Higgs detection at the LHC would probably be
impossible.  We will discuss this more in section~\ref{colliders}.

\begin{figure}[p]
\begin{center}
\epsfig{file=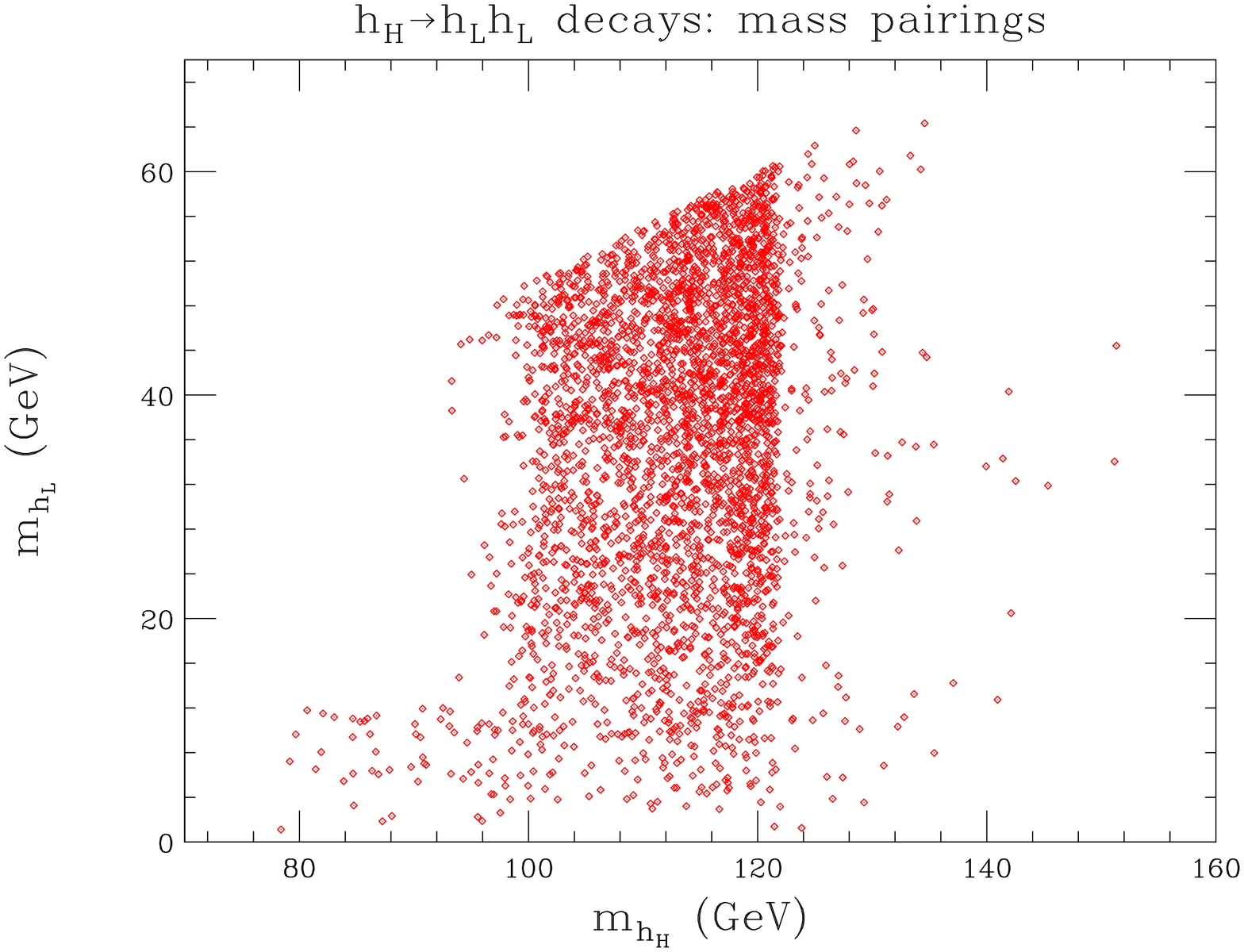,angle=90,width=9.5cm}
\epsfig{file=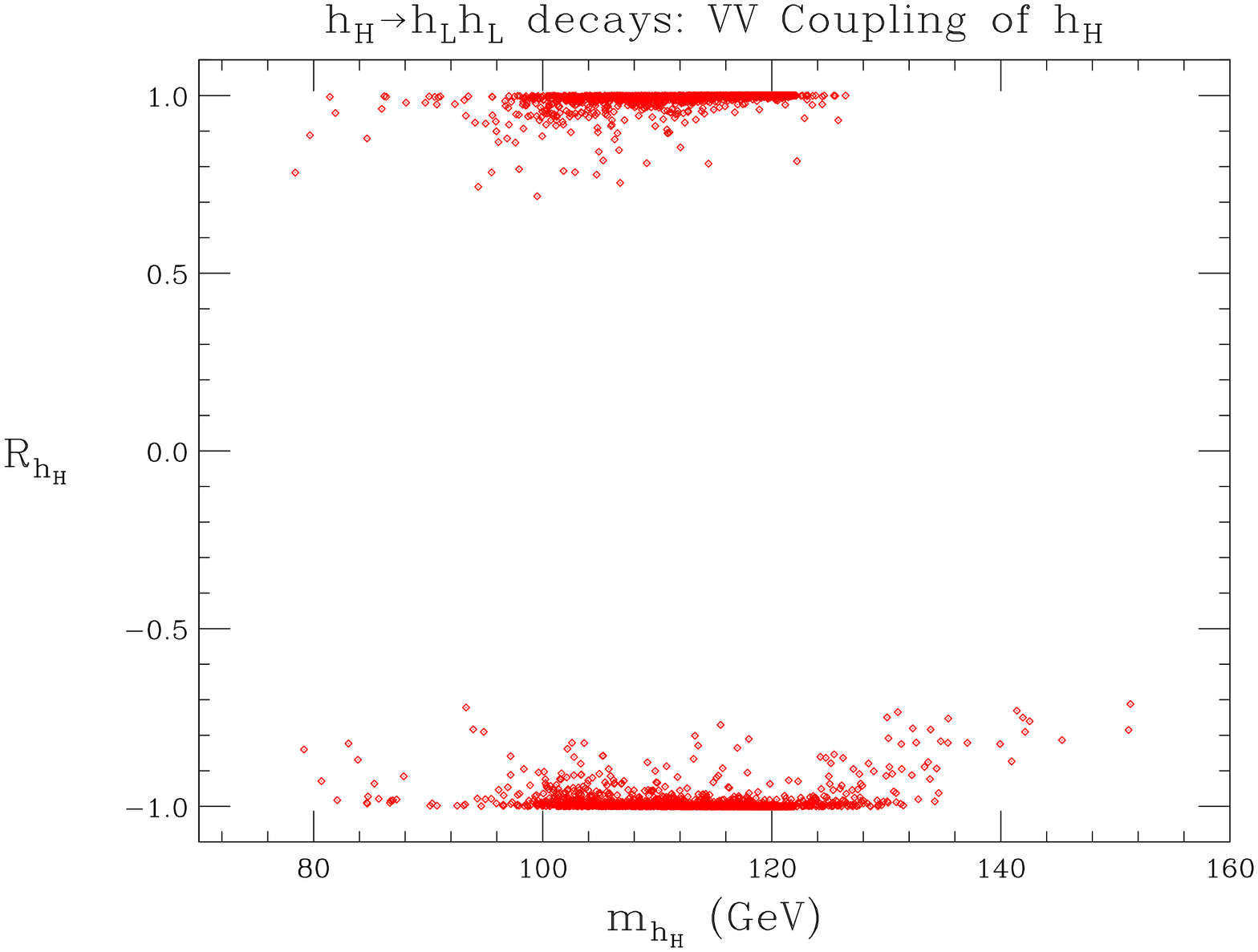,angle=90,width=9.5cm}
\epsfig{file=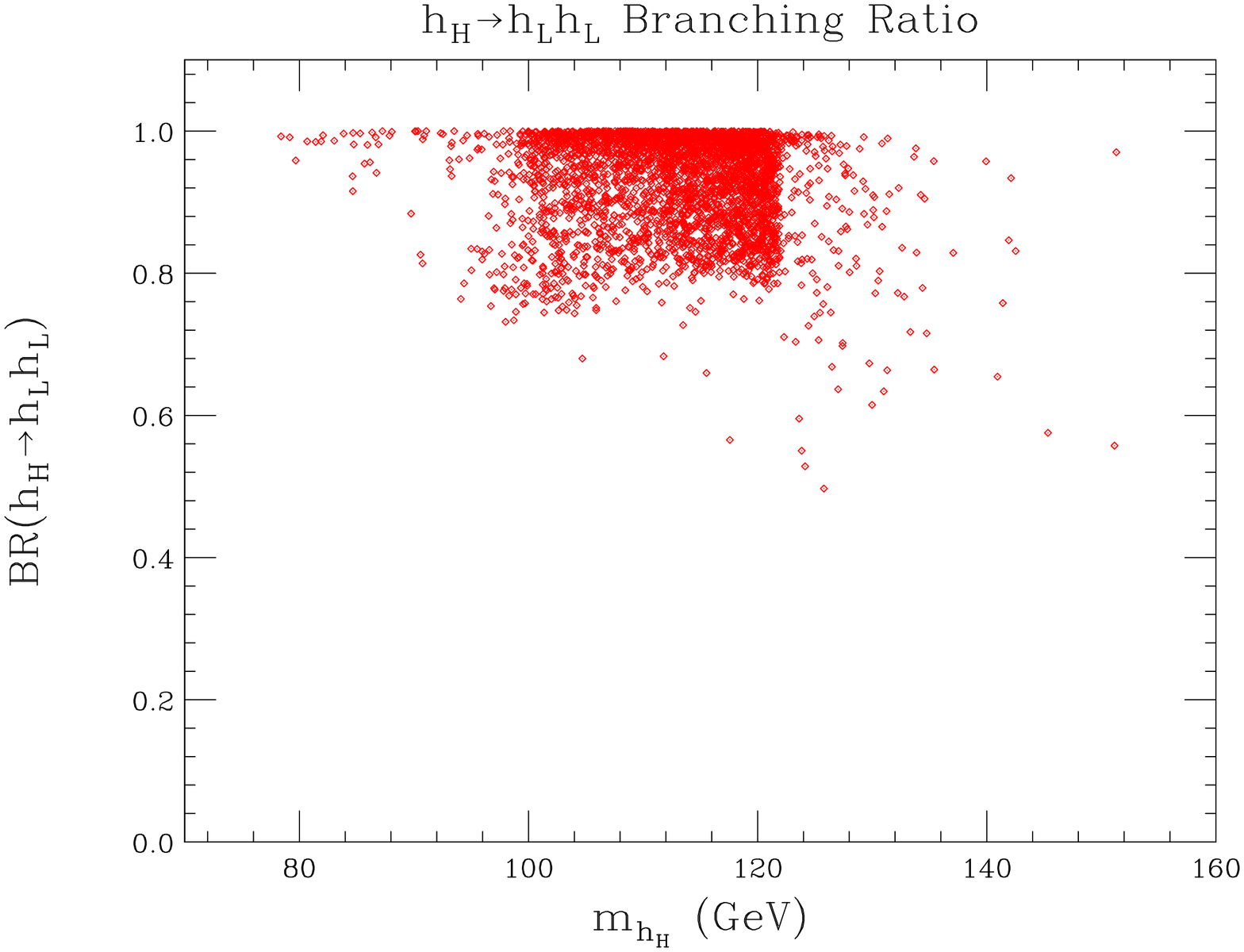,angle=90,width=9.5cm}
\caption{The top, middle and bottom plots give
a scatter plot of $\mhl$, $R_{\hh}$ and
  $\br(\hh\to\hl\hl)$, respectively, 
versus $\mhh$ for 
the $3480$ sample scan points with $\hh\to \hl\hl$ decays.
}
\label{mHvsmL}
\end{center}
\end{figure}

The distribution of the mass of the heavier SM-like Higgs ($\hh$,
where $\hh=h_1$ or $h_2$) as compared to the mass of the lighter Higgs
($\hl$, where $\hl=a_1$ or $h_1$) appearing in the $\hh\to \hl\hl$
decay for the $3480$ points from the scan described above is
given in the top plot of fig.~\ref{mHvsmL}.  We see that the SM-like
parent Higgs mass $m_{\hh}$ lies in the $[75\gev,155\gev]$ interval
while daughter Higgs masses $m_{\hl}$ range from near $0$ up to close
to $m_{\hh}/2$. The middle plot of fig.~\ref{mHvsmL} shows that the
parent $\hh$ always has fairly SM-like coupling $|R_{\hh}|\geq 0.7$ to
vector bosons; for most of the points, $|R_{\hh}|$ is quite close to
$1$.  The bottom plot of fig.~\ref{mHvsmL} shows that for these points
$\br(\hh\to\hl\hl)$ is always substantial.  The importance of the
$WW\to \hh \to \hl\hl$ discovery mode is thus evident.

Out of the above $3480$ points, we have selected eight benchmark points,
the properties of which are displayed in tables~\ref{tpoints} and
\ref{badpoints}, that illustrate the cases where LHC detection of the
NMSSM Higgs bosons in the standard modes 1) -- 9) would not be
possible.  The first five are such that the $WW\to \hh\to \hl\hl\to
jj\tauptaum$ detection mode might be effective.  
Points 6, 7 and 8 are chosen to illustrate
cases where the $\hl$ appearing in the 
final state does not decay to either $b\anti b$
or $\tauptaum$, implying that
the $WW\to \hh\to \hl\hl\to jj\tauptaum$ potential detection mode
would not be useful.

We now discuss in more detail the characteristics of these eight
benchmark points.
\bit
\item
 Points 1, 2 and 3 are designed to illustrate
$h_1\to a_1a_1$ decay cases for a selection of possible $h_1$ and
$a_1$ masses. \\
Point 1 is in the low-mass tail of the $\hh$ mass
distribution (see fig.~\ref{mHvsmL}) at $90\gev$ (although $m_{\hh}$ 
as lows a $80\gev$
is possible).  For point 1, $\mai$ is below the $b\anti b$ threshold so
that the main $\ai$ decay is to $\tauptaum$ or $jj$. \\
Point 2 and point 3 are at
the two extremes of the central bulk of the $\hh$ mass 
distribution of fig.~\ref{mHvsmL} with
$m_{\hh}=100\gev$ and $120\gev$, respectively.  
For these latter two points $\mai$
is $20\gev$ or $30\gev$;  $\ai\to b\anti b$ 
and $\ai\to \tauptaum$ decays will be dominant and in the
usual ratio.
\item
Point 4 is such
that the $\hi$ and $\hii$ (with masses $\mhi=97\gev$ and $\mhii=150\gev$)
share  the $WW/ZZ$ coupling strength squared and both
decay to $\ai\ai$. The $\ai$ decays to $b\anti b$
and $\tauptaum$ in the usual ratio.
Note that this point is an example for which 
$\mai$ is fairly large ($\mai=45\gev$).
\item
Point 5 illustrates a case in which it is the $\hii$
that is SM-like and it decays to $\hi\hi$. The $\hi\to b\anti b$ 
and $\hi\to\tauptaum$ decays are the dominant ones
and are in the usual ratio.  
Although $\mhi$ is rather small
in this case, it would not have been seen at LEP due to its singlet
nature.  
Nonetheless, $BR(\hii\to\hi\hi)$ is large due to the new
trilinear NMSSM couplings.  
\item
For point 6, the $\hi$ is SM-like and decays via $\hi\to\ai\ai$, 
but $\ai\to \gam\gam$ is dominant due to the singlet nature of $\ai$.\
The $4\gam$ final state would provide a highly distinctive signal
that should be easily seen at the LHC~\cite{Dobrescu:2000jt}.
\item
Point 7 illustrates a case in which the $\hii$ is SM-like
and decays via $\hii\to\hi\hi$. The new feature compared to
point 5 is that the $\hi$ has reduced coupling to
$b\anti b$ and $\tauptaum$ due to the fact
that parameters are such that $\hi$ is almost
entirely $H_u$ in nature~\cite{Miller:2004uh}.
\footnote{A continuum
  of points of this type was 
discussed in ref.~\cite{Ellwanger:2004xm}.}
Obviously, the  $WW\to\hii\to jj\tauptaum$ 
 mode would not be relevant for this type of scenario.
 We do not think
that the resulting $h_2\to 4j$ signal could be isolated from
backgrounds. 
\item
Point 8 illustrates a case in which
the $\hi$ is SM-like and decays via $\hi\to\ai\ai$. It differs
from earlier such points in that the 
$\ai$ is extremely light and decays
mainly to $jj$ ($j=s,c,g$).  
Like for point 7, the  $WW\to\hi\to jj\tauptaum$ detection channel
would not be relevant.  We would need to isolate a $\hi\to 4j$
signal within a large QCD background.  We do not believe this
will be possible, especially given that each of the pairs of jets will
have small mass and large boost, making separation of the two jets
within each pair very problematical.\\
This $\ai$ would not have been seen at
LEP in the $\hi\ai$ mode for several reasons (for details
see the references and discussions in~\cite{Ellwanger:2004xm}):
\smallskip
\ben
\item
First, the $Z\hi\ai$ coupling is very small because of
the very SM-like nature of
the $\hi$. 
\item
Second, $\mai$ is below the
threshold of any existing study of the $ha$
type of mode at LEP. 
\item
Third, the existing $Z\hi\to Z\ai\ai$ limits (from OPAL)
for the case where
$\ai\to jj$  do not cover the mass regions corresponding
to the values of either $\mhi$ or $\mai$.
In fact, they extend only up to 
$\mhi\sim 90\gev$ and in any case do not
apply when  $\mai$ is below the $2\gev$ threshold
for their searches. 
\een
\smallskip
Finally, we note that axion searches do not apply since the $\ai$
would not have been invisible in the detector --- it
decays promptly to visible jets.

\eit

\begin{table}[p]
 \begin{center}
 \footnotesize
 \vspace*{-.2in}
 \begin{tabular} {|l|r|r|r|r|r|}
 \hline
 Point Number & 1 & 2 & 3 & 4 & 5 \\
 \hline \hline
 Bare Parameters &\multicolumn{5}{c|}{} \\
 \hline \hline
 $\lambda$               & $ 0.22$ & $  0.4$ & $ 0.22$ & $ 0.67$ & $ 0.56$ \\
 \hline
 $\kappa$                & $ -0.1$ & $-0.35$ & $ 0.59$ & $  0.2$ & $  0.1$ \\
 \hline
 $\tan\beta$             & $   5.$ & $  15.$ & $  7.8$ & $  4.1$ & $  2.5$ \\
 \hline
 $\mu_{\rm eff}$~(GeV)   & $-520.$ & $-160.$ & $ 530.$ & $-200.$ & $-180.$ \\
 \hline
 $A_{\lambda}$~(GeV)     & $-580.$ & $-580.$ & $-920.$ & $-600.$ & $-440.$ \\
 \hline
 $A_{\kappa}$~(GeV)      & $ -2.8$ & $ -8.7$ & $ -2.1$ & $ -30.$ & $ 172.$ \\
 \hline \hline
 CP-even Higgs Boson Masses and Couplings & \multicolumn{5}{c|}{} \\
 \hline \hline
 $m_{h_1}$~(GeV)         & $  90.$ & $ 100.$ & $ 119.$ & $  97.$ & $  40.$ \\
 \hline
 $R_1$                   & $ 0.99$ & $ 0.97$ & $-1.00$ & $ 0.69$ & $ 0.00$ \\
 \hline
 $t_1$                   & $ 0.99$ & $ 0.97$ & $-1.00$ & $ 0.72$ & $ 0.05$ \\
 \hline
 $b_1$                   & $ 1.00$ & $ 0.90$ & $-1.01$ & $ 0.31$ & $-0.35$ \\
 \hline
 $g_1$                   & $ 0.99$ & $ 0.97$ & $ 1.00$ & $ 0.74$ & $ 0.15$ \\
 \hline
 $\gamma_1$              & $ 0.99$ & $ 0.99$ & $ 1.00$ & $ 0.78$ & $ 0.11$ \\
 \hline
 $\br(h_1\to b\anti b)$  & $ 0.08$ & $ 0.02$ & $ 0.01$ & $ 0.01$ & $ 0.93$ \\
 \hline
 $\br(h_1\to\tauptaum)$  & $ 0.01$ & $ 0.00$ & $ 0.00$ & $ 0.00$ & $ 0.07$ \\
 \hline
 $\br(h_1\to a_1a_1)$    & $ 0.91$ & $ 0.97$ & $ 0.99$ & $ 0.99$ & $ 0.00$ \\
 \hline \hline 
 $m_{h_2}$~(GeV)         & $ 479.$ & $ 288.$ & $1431.$ & $ 150.$ & $ 125.$ \\
 \hline
 $R_2$                   & $ 0.16$ & $ 0.26$ & $ 0.00$ & $ 0.72$ & $-1.00$ \\
 \hline
 $t_2$                   & $ 0.16$ & $ 0.26$ & $ 0.13$ & $ 0.70$ & $-1.00$ \\
 \hline
 $b_2$                   & $ 0.19$ & $ 0.57$ & $ -7.8$ & $ 1.10$ & $-1.03$ \\
 \hline
 $g_2$                   & $ 0.16$ & $ 0.26$ & $ 0.12$ & $ 0.69$ & $ 0.99$ \\
 \hline
 $\gamma_2$              & $ 0.19$ & $ 0.24$ & $ 0.08$ & $ 0.65$ & $ 0.99$ \\
 \hline
 $\br(h_2\to a_1a_1)$    & $ 0.04$ & $ 0.44$ & $ 0.00$ & $ 0.97$ & $ 0.00$ \\
 \hline
 $\br(h_2\to h_1h_1)$    & $ 0.25$ & $ 0.21$ & $ 0.00$ & $ 0.00$ & $ 0.92$ \\
 \hline \hline 
 $m_{h_3}$~(GeV)         & $ 952.$ & $1016.$ & $2842.$ & $ 753.$ & $ 495.$ \\
 \hline \hline
 CP-odd Higgs Boson Masses and Couplings &\multicolumn{5}{c|}{} \\
 \hline \hline
 $m_{a_1}$~(GeV)         & $  10.$ & $  20.$ & $  31.$ & $  45.$ & $ 144.$ \\
 \hline
 $t_1'$                  & $-0.01$ & $ 0.00$ & $-0.01$ & $-0.02$ & $-0.06$ \\
 \hline
 $b_1'$                  & $-0.22$ & $-0.85$ & $-0.53$ & $-0.40$ & $-0.40$ \\
 \hline
 $g_1'$                  & $ 0.15$ & $ 0.48$ & $ 0.19$ & $ 0.08$ & $ 0.06$ \\
 \hline
 $\gamma_1'$             & $ 0.12$ & $ 0.11$ & $ 0.15$ & $ 0.49$ & $ 0.61$ \\
 \hline
 $\br(a_1\to b\anti b)$  & $ 0.00$ & $ 0.94$ & $ 0.93$ & $ 0.93$ & $ 0.85$ \\
 \hline
 $\br(a_1\to \tauptaum)$ & $ 0.83$ & $ 0.06$ & $ 0.07$ & $ 0.07$ & $ 0.08$ \\
 \hline
 $\br(a_1\to jj)$        & $ 0.17$ & $ 0.00$ & $ 0.00$ & $ 0.00$ & $ 0.01$ \\
 \hline \hline
 $m_{a_2}$~(GeV)         & $ 952.$ & $1018.$ & $1434.$ & $ 750.$ & $ 495.$ \\
 \hline \hline
 Charged Higgs Boson Mass& \multicolumn{5}{c|}{} \\
 \hline \hline
 $m_{h^\pm}$~(GeV)       & $ 954.$ & $1017.$ & $1432.$ & $ 742.$ & $ 487.$ \\
 \hline \hline
 LSP Mass                & \multicolumn{5}{c|}{} \\
 \hline \hline
 $m_{\cnone}$            & $ 453.$ & $ 136.$ & $ 476.$ & $ 113.$ & $  82.$ \\
 \hline \hline
 Most Visible of the LHC Processes 1)-9)
                         & 2($h_1$)& 5($h_2$)& 2($h_1$)& 5($h_2$)& 2($h_2$) \\
 \hline
 $N_{SD}=S/\sqrt B$ of this process at $L=$300~${\rm fb}^{-1}$
                         & $  1.6$ & $  0.7$ & $  0.3$ & $  0.5$ & $  2.0$ \\
 \hline
 \end{tabular}
 \end{center}
\caption{\label{tpoints}\footnotesize
Properties of five scenarios for which LHC Higgs detection would only be
possible in the $WW\to h_{1,2}\to a_1a_1\to jj\tauptaum$  or $WW\to h_2\to
h_1h_1\to jj\tauptaum$ mode.   The quantities $R_i$, $t_i$, $b_i$, $g_i$, $\gam_i$, 
$t_i'$, $b_i'$, $g_i'$ and $\gam_i'$ were defined in the caption of table~\ref{higgsprop}.
Important absolute branching ratios are displayed. Only the masses of the heavy
$h_3$, $a_2$ and $h^{\pm}$ are given. The mass of the lightest neutralino (LSP)
is also given. The second-to-last row gives the channel and
Higgs boson yielding the largest $N_{SD}=S/\sqrt B$ in channels 1) -- 9).  The following
row gives the corresponding $N_{SD}$ for $L=300\fbi$.
}
\end{table}

\begin{table}[p]
 \begin{center}
 \footnotesize
 \vspace*{-.2in}
 \begin{tabular} {|l|r|r|r|}
 \hline
 Point Number & 6 & 7 & 8 \\
 \hline \hline
 Bare Parameters & \multicolumn{3}{c|}{} \\
 \hline \hline
 $\lambda$                & $ 0.39$ & $  0.5$ & $ 0.27$ \\
 \hline
 $\kappa$                 & $ 0.18$ & $-0.15$ & $ 0.15$ \\
 \hline
 $\tan\beta$              & $  3.5$ & $  3.5$ & $  2.9$ \\
 \hline
 $\mu_{\rm eff}$          & $-245.$ & $ 200.$ & $-753.$ \\
 \hline
 $A_{\lambda}$            & $-230.$ & $ 780.$ & $ 312.$ \\
 \hline
 $A_{\kappa}$             & $  -5.$ & $ 230.$ & $  8.4$ \\
 \hline \hline
 CP-even Higgs Boson Masses and Couplings & \multicolumn{3}{c|}{} \\
 \hline \hline
 $m_{h_1}$~(GeV)          & $  94.$ & $  57.$ & $  95.$ \\
 \hline
 $R_1$                    & $ 0.94$ & $-0.28$ & $ 1.00$ \\
 \hline
 $t_1$                    & $ 0.95$ & $-0.30$ & $ 0.99$ \\
 \hline
 $b_1$                    & $ 0.89$ & $ 0.01$ & $ 1.05$ \\
 \hline
 $g_1$                    & $ 0.95$ & $ 0.33$ & $ 0.99$ \\
 \hline
 $\gamma_1$               & $ 0.96$ & $ 0.37$ & $ 1.00$ \\
 \hline
 $\br(h_1\to jj)$         & $ 0.01$ & $ 0.93$ & $ 0.00$ \\
 \hline
 $\br(h_1\to a_1a_1)$     & $ 0.94$ & $ 0.00$ & $ 1.00$ \\
 \hline \hline 
 $m_{h_2}$~(GeV)          & $ 239.$ & $ 125.$ & $ 483.$ \\
 \hline
 $R_2$                    & $ 0.33$ & $-0.96$ & $-0.01$ \\
 \hline
 $t_2$                    & $ 0.30$ & $-0.95$ & $-0.36$ \\
 \hline
 $b_2$                    & $ 0.67$ & $-1.07$ & $ 2.84$ \\
 \hline
 $g_2$                    & $ 0.29$ & $ 0.95$ & $ 0.37$ \\
 \hline
 $\gamma_2$               & $ 0.30$ & $ 0.94$ & $ 0.68$ \\
 \hline
 $\br(h_2\to h_1h_1)$     & $ 0.32$ & $ 0.93$ & $ 0.01$ \\
 \hline \hline 
 $m_{h_3}$~(GeV)          & $ 562.$ & $ 731.$ & $ 821.$ \\
 \hline \hline
 CP-odd Higgs Boson Masses and Couplings & \multicolumn{3}{c|}{} \\
 \hline \hline
 $m_{a_1}$~(GeV)          & $  40.$ & $ 188.$ & $   1.$ \\
 \hline
 $t_1'$                   & $ 0.00$ & $ 0.04$ & $ 0.08$ \\
 \hline
 $b_1'$                   & $ 0.00$ & $ 0.53$ & $ 0.62$ \\
 \hline
 $g_1'$                   & $ 0.00$ & $ 0.04$ & $ 0.36$ \\
 \hline
 $\gamma_1'$              & $ 0.47$ & $ 0.31$ & $ 0.39$ \\
 \hline
 $\br(a_1\to jj)$         & $ 0.00$ & $ 0.00$ & $ 0.95$ \\
 \hline
 $\br(a_1\to \mu\mu)$     & $ 0.00$ & $ 0.00$ & $ 0.05$ \\
 \hline
 $\br(a_1\to\gam\gam)$    & $ 0.98$ & $ 0.00$ & $ 0.00$ \\
 \hline
 $\br(a_1\to\cnone\cnone)$& $ 0.00$ & $ 0.99$ & $ 0.00$ \\
 \hline \hline
 $m_{a_2}$~(GeV)          & $ 558.$ & $ 736.$ & $ 493.$ \\
 \hline \hline
 Charged Higgs Boson Mass & \multicolumn{3}{c|}{} \\
 \hline \hline
 $m_{h^\pm}$~(GeV)        & $ 560.$ & $ 727.$ & $ 485.$ \\
 \hline \hline
 LSP Mass                 & \multicolumn{3}{c|}{} \\
 \hline \hline
 $m_{\cnone}$               & $ 211.$ & $  81.$ & $ 500.$ \\
 \hline \hline
 Most Visible of the LHC Processes 1)-9)
                          & 5($h_2$)& 2($h_2$)& 5($h_3$)\\
 \hline
 $N_{SD}=S/\sqrt B$ of  this process at $L=$300~${\rm fb}^{-1}$
                          & $  1.5$ & $  1.3$ & $  0.1$ \\
 \hline
 \end{tabular}
 \end{center}
\caption{\label{badpoints}\footnotesize
Properties of three representative scenarios for which LHC Higgs detection would
not even be possible in the $WW\to h_{1,2}\to a_1a_1\to jj\tauptaum$ or $WW\to
h_2\to h_1h_1\to jj\tauptaum$ modes.
}
\end{table}

\section{Collider Implications}
\label{colliders}

In the previous section, we have established
the probable importance and possible necessity of
detecting a fairly SM-like, relatively light
 Higgs boson, $\hh$, in a Higgs-pair
decay mode, $\hh\to\hl\hl$. In this section, we discuss
possible ways in which such detection might be possible at
various different kinds of colliders, with some emphasis 
on the LHC. The best means for such detection at hadron
colliders will depend strongly upon the $\hl$ decay channels.
Detection of the $\hh$ in the Higgs-pair final state
at an $\epem$ or $\gam\gam$ collider will be less dependent
upon precisely how the $\hl$ decays.

\noindent\underline{The LHC}

At the LHC it will presumably be
highly advantageous to use $WW$ fusion production for the $\hh$.
Not only is the associated production cross section quite competitive
with other production mechanisms for $\hh$ in the $\mhh\in[80\gev,150\gev]$
mass range of relevance, but also 
the ability to tag the spectator jets will certainly
make backgrounds much more manageable. 

When $\mhl>2m_b$, we advocate employing the $\hl\hl\to b\anti b \tauptaum$
final state. In this final state an approximate mass
for the $\hh$ can be computed using the visible
particles in the final state to compute an effective mass,
$\mjjtautau$. In addition, restrictions
on the visible mass of each $\hl$ can also be imposed.
(In the analysis, one will need to choose a hypothetical value
for $\mhl$ and examine the $\mjjtautau$ mass distribution for
a peak. This process will have to be repeated for all possible $\mhl$
values.  Only the choice with $\mhl$ near the actual value would
reveal a peak in $\mjjtautau$.) 
For the $b\anti b\tauptaum$ final state, the two main backgrounds
appear to be: (i) $gg\to t\anti t\to b\anti b \wp \wm\to b\anti b
\tauptaum+\etmiss$, in association with forward and backward jet radiation
and (ii) Drell-Yan $\tauptaum+jets$
production. Monte Carlo simulations performed to date 
show that $b$-tagging does not seem to be necessary 
to overcome the a priori large Drell-Yan
$\tauptaum+jets$ background.  It is eliminated
by stringent cuts for finding the highly energetic forward / backward
jets characteristic of the $WW$ fusion  process. 
To the extent that the main background 
will then come from $t\anti t$ production, it
is not useful to specifically $b$-tag the $b$ jets since
the $t\anti t$ background will also contain $b$ jets.
Thus, it is appropriate to
focus on a generic $WW\to jj\tauptaum+\etmiss$ final state.
Such a final state can be experimentally isolated with high efficiency
by identifying two $\tau$'s from one $a_1$ using the  leptonic decay
modes for both $\tau$'s 
while requiring two ($b$) jets from the other $a_1$.
All these particles should be required to be quite central. 

If $\mhl<2m_b$, but above $2\mtau$,
the dominant final state is $\hl\hl\to\tauptaum\tauptaum$.  However,
an effective $\hh$ mass is very difficult to reconstruct in this
channel.  Previous work suggests that it will be best
to employ the $\hl\hl\to jj\tauptaum$ final state (which typically
has a small but usable branching ratio) where
one of the $\hl$'s decays to $jj=c\anti c$, $s\anti s$ or $gg$.
This is extracted experimentally by again 
identifying two $\tau$'s in their leptonic decay modes
and two jets.

Preliminary
simulations for the $jj\tauptaum$ signal have appeared 
in~\cite{Ellwanger:2003jt,Ellwanger:2004gz} for a few representative
benchmark points (different from those appearing in the tables
of this paper). Aside from imposing stringent
forward / backward jet tagging cuts to eliminate the Drell-Yan
$\tauptaum+jets$ background, it was required that the two additional 
jets (from one of the $\hl$'s) and the two opposite sign
central leptons ($\ell=e,\mu$) coming from the 
the $\tau^+\tau^-$ emerging from the decay of the other
$\hl$ all be quite central. Additional observations are the following:
\bit
\item In the case of $\mhl>2m_b$, the
$jj$ mass is fairly high, efficiencies for identifying such  $j$'s
were found to be high and the momenta of each $j$ were relatively
well determined.  The latter implies that 
the $\tauptaum$ mass can be reconstructed by assuming that all missing
transverse momentum is to be associated with the neutrinos
in the $\tauptaum\to \ell^+\ell^- +\nu's$ decays.  Efficiencies
for the overall reconstruction of this kind of event are therefore
reasonably high.  

\item In the case of $2m_\tau<\mhl<2m_b$, 
the primary source of the $jj$
is from $c\anti c+s\anti s +gg$ decays of one of the $\hl$'s
(neglecting the inefficiently tagged and poorly reconstructed
contribution coming from $\hl\to\tauptaum$
with two $\tau\to j+\etmiss$ decays, where the $j$ is a
single pion or similar hadronic resonance).  
However, the $jj$ pair mass is quite low ($<2\mb$) and separate
identification of the two jets was found to be rather inefficient.

For this case, the relevant
final state branching ratio is $\br(\hl\hl\to jj\tauptaum)=2\times\br(\hl\to c\anti c +s\anti
s+gg)\ \br(\hl\to\tauptaum)$, which is similar in size (typically) for
$2m_\tau<\mhl<2\mb$ to what one finds when $\mhl>2\mb$ for $\br(\hl\hl\to
b\anti b\tauptaum)$.
\eit

We reiterate that since the $\hl$ will not 
have been detected previously, 
we must assume a value for $\mhl$ to perform the analysis. 
We then look among the central jets for the combination with invariant mass
$M_{jj}$ closest to $\mhl$ (no $b$-tagging is enforced, $b$'s
are identified as non-forward/backward jets).  
We then compute the $\mjjtautau$ invariant mass using the
four reconstructed four-momenta for the two $j$'s and two $\tau$'s
and look for a bump in the distribution. This process is repeated
for densely spaced $\mhl$ values and we
look for the $\mhl$ choice that produces
the best signal. 

In our earlier simulations of points with $\mhh\in[115\gev,130\gev]$, 
the typical result found (for 
the assumed $\mhl$ chosen to agree with the actual $\mhl$)
was a sizable bump in $\mjjtautau$ coming from the signal in the
$\mjjtautau$ range from roughly $40\gev$ to $130\gev$.
The $t\anti t$ background produces a huge peak in $\mjjtautau$
at high mass with a rapidly falling tail 
in the $\mjjtautau\lsim 120\gev$ region. The crucial issue is
the precise shape of this tail and exactly how it extends into
the low-$\mjjtautau$ region where the signal peak resides.
The early results of ~\cite{Ellwanger:2003jt,Ellwanger:2004gz},
based on UA1 detector resolutions and efficiencies, found that
the low-$\mjjtautau$ tail from $t\anti t$ does not overlap significantly
the signal bump, which (for $L=300\fbi$) 
typically contained between 500 and 2000 events
in  $\mjjtautau\in[40\gev,130\gev]$. More recently, members of the
ATLAS collaboration~\cite{atlascol} have examined
the $\mjjtautau$ signal using the ATLAS
simulation programs and somewhat different cuts.  They find that
the $t\anti t$ background extends over the full range where the
signal resides.  
This question is now being examined using full simulations
in the context of the ATLAS detector and improved
cuts~\cite{atlascol,preparation}. We are unable at
this time to say whether or not the signal will emerge above
the background in a statistically significant and reliable way.

In principle, one could explore final states other than the
$\hl\hl\to jj\tauptaum$ mode. However, all other channels will be much more
problematical at the LHC. A $4b$-signal would be present
for $m_{\hl}>2\mb$ but would
be burdened by a large QCD background even after
implementing $b$-tagging.  Meanwhile, for $2\mtau<m_{\hl}<2\mb$
the $4\tau$-channel would typically have a large branching ratio
and we could look for it in the mode where all $\tau$'s decay
leptonically. However, in this mode it would not be possible to
reconstruct the $\hh$ resonance mass and backgrounds would be large.  

It should be clear that without the ability to tag one $\tauptaum$
pair as part of the $\hh$ reconstruction process, the background
would be much larger.  Thus, it is only for  $\mhl>2\mtau$ 
that there is a possibility to isolate the $WW\to\hh\to\hl\hl$ signal.
We are very pessimistic regarding isolating a significant signal
in a $4j$ final state as appropriate when $\mhl<2\mtau$ or in
those special cases where $\mhl>2\mtau$ but $\hl\to jj$ ($j=c,s,g$)
decays are dominant.  

\noindent\underline{The Tevatron}

At the Tevatron, $WW\to \hh$ production has a rather low cross
section. Only the $gg\to \hh$ cross section is sizable (of order
$1\pb$ for a SM-like $\hh$ with $\mhh\sim 100\gev$).  For the case of
$\mhl>2m_b$, one would again employ the $\hl\hl\to b\anti b\tauptaum$
final state.  However, forward / backward jet tagging could no longer
be used to reduce the Drell-Yan $\tauptaum+jets$ background without
also severely affecting the $gg$ fusion signal.  The best means for
discriminating against the DY background would probably be to use
$b$-tagging.  Of course, this will not reduce the dominant $t\anti t$
background relative to the Higgs signal.  Detailed simulations will be
required to see if a signal can be extracted..  A
group~\cite{davisgroup} including CDF experimentalists is working on
simulating typical cases with $2\mtau<\mhl<2m_b$. They are currently
focusing on the $\tauptaum\tauptaum$ final state where one $\tau$ is
identified through its $\tau\to \mu+\etmiss$ decay and all other
$\tau$'s are identified using either an isolated hadronic decay
signature or a lepton decay.

\noindent\underline{An $\epem$ linear collider}

At an $\epem$ collider, it will be possible to detect any relatively
light Higgs boson with substantial $ZZ$ coupling using the $\epem\to
Z^*\to ZX$ final state and searching for the prominent peak in $\mx$
that would arise if a Higgs boson is present.  This technique is
completely independent of the Higgs decay mode. Once a peak is found,
it would be straightforward to isolate the $\hh\to\hl\hl$ final state
in various $\hl$ decay modes and check if the $\hl$ branching ratios
are consistent with expectations for a light $\hi$ or $\ai$ of the
observed mass. In addition, $\mhh$, $\mhl$, $g_{ZZ\hh}$, and
$g_{\hh\hl\hl}$ will all be measured with considerable precision.
This would allow precision tests of the NMSSM model structure,
especially if part of the supersymmetric particle spectrum is also
accessible.

\noindent\underline{A $\gam\gam$ collider}

Another facility of particular interest for the kind of scenario
presented here will be a $\gam\gam$ collider.  Since the $\hh$
is typically quite SM-like, it will have a very substantial production
rate in $\gam\gam$ collisions. A recent study~\cite{Gunion:2004si}
shows that a very substantial signal for
the $\gam\gam\to\hh\to\hl\hl$ process will be present
above a very small background (after appropriate simple cuts)
in the main $\hl\hl\to b\anti b b\anti b$ and $\hl\hl\to b\anti b \tauptaum$ 
final states. Excellent determinations of both $\mhh$ and $\mhl$ will be
possible and the $\gam\gam$ coupling of the $\hh$ will be very
precisely determined, as will the $\hh\to\hl\hl$ coupling strength.
\pagebreak

\section{Conclusions}
\label{conclusions}

In summary, we have explored the NMSSM model parameter space, looking
for Higgs sector scenarios consistent with LEP exclusions that might
be unexpectedly difficult to probe at the LHC in the conventional
modes that have been explored for the SM and the MSSM.  We have found
that generic points in NMSSM parameter space are such that
Higgs-to-Higgs decays are present. This is a crucial
issue since hadron collider signals for Higgs bosons
decaying to other Higgs bosons will typically be much more difficult 
to extract in the presence of backgrounds than signals in the
conventional modes studied for the SM/MSSM scenarios.

In section~\ref{nohiggstohiggs}, we considered NMSSM parameter points
for which decays of neutral Higgs bosons to other Higgs bosons
were not present.  For this small fraction of parameter space,
we are able to show that the conventional SM/MSSM Higgs boson discovery
modes 1) -- 9) (as listed at the beginning
of the section~\ref{nohiggstohiggs}) 
are sufficient (assuming $L=300\fbi$) to guarantee that
at least one NMSSM Higgs boson will be detectable at the $\geq
5\sigma$ level at the LHC. The worst point yielded 
signals between $6\sigma$ and $7\sigma$
in several of the standard modes for several
different Higgs bosons. However, the limited statistical 
significance for these signals means
that if the effective integrated luminosity falls below $L\sim
100\fbi$ or if backgrounds are larger than found in the 
simulations, then this `no-lose' theorem would fail. In any case,
for the most difficult no-Higgs-to-Higgs points Higgs discovery will
not be easy or quick --- considerable thoroughness and patience will
be required. The interplay between different detection channels
and different Higgs states will be crucial.  Good statistical
significance might only be achieved by combining a number of channels.

The vast bulk of physically acceptable NMSSM parameter
choices are  such that Higgs-to-Higgs decays are present. The focus
of this paper has been to isolate those cases where these decays 
reduce statistical significances in all the standard modes to 
a level such that there is no $\geq 5\sigma$ level signal in
any standard mode for any Higgs boson. In section~\ref{higgstohiggs},
we presented eight sample points for which all the standard
modes had very low statistical significance and detection 
of a SM-like $\hh$ decaying to a pair of lighter $\hl$'s would
provide the only possible signal. Five of the sample points
were such that the final state of interest would be
$\hl\hl\to jj \tauptaum$ (for $2m_\tau<\mhl<2m_b$) or $\hl\hl\to
b\anti b\tauptaum$ (for $2m_b<\mhl$). We noted the potential 
importance of the LHC channel $WW\to \hh\to \hl\hl\to jj\tauptaum$
for detecting a Higgs signal in these cases.
Two of the other three points
were such that only $\hl\to jj$ ($j=c,s,g$) decays were present
and the last point was such that $\hl\to \gam\gam$.  In the latter case,
the $4\gam$ final state would provide a very clean LHC signal.
In the former two cases, we are unable to envisage a technique
for discovering any of the Higgs bosons at the LHC.

In section~\ref{colliders}, we pursued further the issue of Higgs
detection at the LHC for cases like those of the first five
sample points noted above.  We discussed the nature
of and techniques for extracting the  $WW\to \hh\to \hl\hl\to
jj\tauptaum$ LHC signal.  As noted there, this signal is being actively
worked on in collaboration with members of ATLAS.  We also presented a
brief summary of the Higgs boson signal at the Tevatron
based on $gg\to \hh$ fusion.  Finally, we summarized why it is
that at an $\epem$ collider or $\gam\gam$ collider it will be
far easier to detect $\hh$ production followed by $\hh\to\hl\hl$
decay than at a hadron collider. For instance, at the ILC, discovery
of a light SM-like $h$ is guaranteed to be
possible in the $Z\hh$ final state using the decay-independent recoil
mass technique~\cite{Gunion:2003fd}.   

Regarding the scenarios for which only the $WW\to \hh\to \hl\hl\to
jj\tauptaum$ channel might provide a signal at the LHC, we note that
the main issue will be whether the background from $t\anti t$
production (which we believe is the primary background after
appropriate cuts requiring highly energetic forward / backward jets to
eliminate the DY $\tauptaum+jets$ background) will extend to low
values of the reconstructed $\mjjtautau$ mass where the signal
resides. To answer this question requires a very full simulation.
However, it is essential that the ATLAS and CMS groups attack this
problem vigorously since, in the worst case scenarios, this signal
will be the only evidence for Higgs bosons at the LHC. Once the LHC is
operating, the $t\anti t$ background can be more completely modeled
and the significance of any enhancement observed in the $\mjjtautau$
distribution more reliably assessed.  However, even if a fully
trustworthy signal is seen at the LHC, a future ILC will probably be
essential in order to confirm that the enhancement seen at the LHC
really does correspond to a Higgs boson.

We should also note that, for parameter space points of
the type we have discussed here, detection of any of the other NMSSM
Higgs bosons is likely to be impossible at the LHC and is likely to
require an ILC with $\sqrt{s_{e^+e^-}}$ above the relevant thresholds
for $h'a'$ production, where $h'$ and $a'$ are heavy CP-even and
CP-odd Higgs bosons, respectively.

Although the scan results presented here were done for sparticles 
(except possibly the $\cnone$)
that are fairly heavy, we do not believe the results will change
significantly if the sparticles are as low in mass
as current LEP and Tevatron bounds.  
This is because the primary issue is how the SM-like Higgs boson
(which must have mass below roughly $150\gev$ when perturbativity
up to the GUT scale is imposed) decays. Its decays will not
be significantly affected by sparticles with masses even slightly
above current limits.

At the LHC, if SUSY is
discovered and $WW\to WW$ scattering is found to
be perturbative at $WW$ energies of 1 TeV (and
higher), and yet no Higgs bosons are detected in
the standard  modes, a careful search for the
signal we have considered should have a high
priority.  

Finally, we should remark that the
$\hh\to \hl\hl$ search channel considered here in the
NMSSM framework is also highly relevant for a
general two-Higgs-doublet model, 2HDM.  It is
really quite possible that the most SM-like
CP-even Higgs boson of a 2HDM will decay primarily
to two CP-odd states.  This is possible even if
the CP-even state is quite heavy, unlike the NMSSM
cases considered here. If CP violation is introduced
in the Higgs sector, either at tree-level
or as a result of one-loop corrections, then 
$\hh\to\hl\hl$ decays will generally be present (as, for example,
in the CP-violating MSSM ~\cite{Carena:2002bb}). The 
critical signal will be the same as
that considered here.
\pagebreak

\subsubsection*{Acknowledgments}
JFG is supported by the U.S. Department of Energy and the Davis
Institute for High Energy Physics. 
The authors thank
the France-Berkeley fund for partial support of this research.

We are deeply indebted to our many experimental colleagues who
aided us in obtaining the needed LHC simulation inputs for the
various standard LHC discovery channels considered:  
J. Cammin, V. Drollinger, R. Kinnunen,
K. Lassila-Perini,  A. Nikitenko, and M. Sapinski.
We are also very grateful for the continued collaboration of
S. Baffioni, S. Moretti and D. Zerwas in simulating the Higgs-to-Higgs decay
signals at the LHC and we thank D. Miller for many useful
conversations.


\newpage
\section*{Appendix A: Summary of ATLAS and CMS simulations employed and
rescaling procedures}

We had a large number of experimental simulations available
for each of the standard discovery channels 1) -- 9).
Because of the need to go to $L=300\fbi$ in order to 
achieve a firm no-lose theorem for NMSSM Higgs discovery
in the absence of Higgs-to-Higgs decays, whenever available
we employed results for CMS or ATLAS for $L=100\fbi$
rather than low luminosity, $L=30\fbi$, results.
In some channels, the CMS results indicated greater discovery
potential than ATLAS results and vice versa.
We always employed the best {\it single detector} results.
(That is, we do not double the statistics assuming two detectors.) 
We did not make use of any studies other than those performed by the
ATLAS and CMS detector collaborations.
 We do not
attempt to give all the different simulations considered but
only summarize those we actually used for each of the nine 
standard channels. We apologize in advance for not referencing all
the experimental (and theoretical) studies that we did not end up
using.

To be conservative, we always employed results obtained for the case
where the radiative correction ``$K$ factors'' for the signal and
background were unity: $K_S=1$.  and $K_B=1$. At the LHC, it is almost always the case that the actual $K$
factors for the signal and background (before cuts) for a given
channel are such that $\frac{S}{\sqrt B}$ improves upon their inclusion.
But, using the $K$ factors obtained before cuts is unreliable since
the $K$ factors can easily be sensitive to the cuts and selection
procedures employed by the experimental groups. Eventually, full,
process-specific Monte Carlos will be available at NLO that will allow
$K$ factor evaluation after cuts.  at which time this kind of study
could be repeated in order to see if the radiative corrections have
significant impact.  For a recent summary of LHC radiative corrections
related to Higgs production and decay, see \cite{Assamagan:2004mu} and
references therein.

\begin{table}[h!]
\caption{Resolutions for combining overlapping signals in a given
  channel.
\label{deltatab}}
\smallskip
\begin{center}
\begin{tabular} {|l||c|c|c|c|c|c|c|c|c|}
\hline
Channel Number &  1 & 2 & 3 & 4 & 5 & 6 & 7 & 8 & 9  \\
\hline
$\Delta(i)$   & 0.01 & 0.01 & 0.10 & 0.15 & 0.01 & 0.01 & 0.10 & 0.10 & 0.10  \\
\hline
\end{tabular}
\end{center}
\end{table}

Finally, we must account for the
fact that the different $h_i$ and $a_i$ 
can have a range of different masses, sometimes overlapping, sometimes
not. Thus, signals in a given discovery
channel from different scalars and/or pseudo-scalars can overlap
within the experimental resolution.  In this case, the overlapping
signals should be combined. We
have chosen to combine the scalar and/or pseudo-scalar signals at different
masses following the procedure of ref. \cite{richterwas}, section 5.4,
using a channel-dependent resolution. In particular, we have
chosen to employ (in the notation of \cite{richterwas}) 
$\sigma_m(i)=2\Delta(i) \times m$ ($m$ being the Higgs
mass, and $i$ denoting the channel) with the $\Delta(i)$ values as
given in table~\ref{deltatab}.  A particularly relevant
example is channel 4) (in the sense that there is often overlap between
scalar and pseudo-scalar Higgs boson resonance signals
which individually have a useful level of significance). For channel 4) we
estimated $\Delta=0.15$ from
fig. 19-61 in \cite{unknown:1999fr} at high luminosity and
extrapolated to $m_A\ \lsim\ 150$ GeV.

\bed
\item{Channel 1):} For $gg \to h\to \gam\gam$ we employ $L=100\fbi$
results analogous to those for $L=30\fbi$  contained in fig.~1 of \cite{Kinnunen:2002}.
For our purpose it is crucial to avoid summing over the
$gg\to h\to \gam\gam$  and  $Wh+t\anti t h\to \ell \gam\gam X$
channels.   This is because
production rates in these two channels are scaled differently in the
NMSSM, 
the first being scaled by the factor $g_i^2$ and the second by
a combination of $R_i^2$ and $t_i^2$. 
We thank R. Kinnunen, K. Lassila-Perini and A. Nikitenko for providing
us with this separation in a series of email communications.
For $L=100\fbi$, the resulting $S/\sqrt B$ values 
for the $gg\to \hsm$ fusion process alone are summarized
in table~\ref{chan1} below for the assumption that $K$ factors for
signal and background are both unity: $K_S=K_B=1$.  

The production rates 
in the $gg$ channel must be corrected for non-SM-like
$gg$ couplings of the Higgs bosons. We must also account
for differences in the $\gam\gam$ branching ratio relative
to that of the SM Higgs boson.
Thus, the tabulated entries are to be multiplied
by $g_i^2\br(h_i\to \gam\gam)/\br(\hsm\to \gam\gam)$ 
for the $h_{1,2,3}$
and by $g_i^{\prime \,2}\br(a_i\to\gam\gam)/\br(\hsm\to\gam\gam)$ for the $a_{1,2}$.
The $L=300\fbi$ results are obtained
by scaling the results so obtained by $\left[\frac{300\fbi}{100\fbi}\right]^{1/2}$.
\begin{table}[h!]
\caption{ $gg \to h \to \gamma\gamma$: CMS, $L=100\fbi$, $K_S=K_B=1$ \label{chan1}}
\smallskip
\begin{center}
\begin{tabular} {|l||l|l|l|l|l|l|}
\hline
$m$ [GeV] &  100 & 110 & 120 & 130 & 140 & 150  \\
\hline
$S/\sqrt{B}$   & 4.2 & 6.0 & 6.8 & 8.2 & 7.0 &
5.2  \\
\hline
\end{tabular}
\end{center}
\end{table}
\item{Channel 2):} For $Wh+t\anti t h\to \gam\gam$, we  employ
the CMS $L=100\fbi$ results for this separate channel, 
as provided to us by R. Kinnunen, K. Lassila-Perini and A. Nikitenko.
These are tabulated in table~\ref{chan2}. 
Since the $S/\sqrt B$ in these SM-Higgs simulations came about
50\% from the $Wh$ channel and about 50\% from the $t\anti t h$
channel, we rescale the production rate for this process by
$\half(R_i^2+t_i^2)$ 
for $h_{1,2,3}$ or $\half t_i^{\prime\, 2}$ for the $a_{1,2}$.
Including the correction for the $\gam\gam$ branching ratio, 
the tabulated results are rescaled by
$\half(R_i^2+t_i^2)\br(h_i\to\gam\gam) / \br(\hsm\to\gam\gam)$ for
the $h_{1,2,3}$ and by $\half t_i^{\prime\,
  2}\br(a_i\to\gam\gam) / \br(\hsm\to\gam\gam)$ for the $a_{1,2}$.
The $L=300\fbi$ results are obtained
by scaling the results so obtained by
$\left[\frac{300\fbi}{100\fbi}\right]^{1/2}$.

\begin{table}[h!]
\caption{ $Wh / t\bar{t}h \to \gamma\gamma l$: CMS, $L=100\fbi$,
  $K_S=1$, $K_B=1$ \label{chan2}}
\smallskip
\begin{center}
\begin{tabular} {|l||l|l|l|l|l|l|l|l|l|l|l|} 
\hline
$m$ [GeV] 
     & 80 & 90   & 100  & 110  & 120  & 130  & 140  & 150  \\
\hline
$S/\sqrt{B}$ 
     & 9.4 & 10.6 & 10.9 & 14.8 & 15.7 & 13.2 & 10.4 & 8.2  \\
\hline
\end{tabular}
\end{center}
\end{table}

\item{Channel 3):} For $t\anti t h\to t \anti t b\anti b$,
we employed results supplied by V. Drollinger
based on extension of the work in ref.~\cite{Drollinger:2001}
to the much larger Higgs mass range required for our NMSSM study.
We are very grateful for these additional results, which were
absolutely critical to our study, and for the collaboration
of V. Drollinger in checking the final table~\ref{chan3} below, including:
the extrapolation to $L=100\fbi$; the change
from $K_B=1.9$ to $K_B=1$ (the standard we employ in this paper);
and the removal of the SM result for $\br(\hsm\to b\anti b)$ ---
the results of table~\ref{chan3} are to be multiplied by
$\br(h,a\to b\anti b)$ and not the ratio to the SM
Higgs branching ratio.
V. Drollinger emphasizes that the extrapolation to $L=100\fbi$
has ignored beam pile-up which  might cause some diminution in
$b$-tagging efficiency at the higher $L=100\fbi$ luminosity.
(This will be studied during preparation of the CMS TDR.)
Thus, we have been somewhat cautious in extrapolating table~\ref{chan3}
to the full $L=300\fbi$ luminosity by employing the factor
$\left[\frac{300\fbi}{100\fbi}\right]^{1/4}$. As regards rescaling
this table for the various NMSSM Higgs bosons,
the results given are to be multiplied by 
$t_i^2\br(h_i\to b\anti b)$ for $h_{1,2,3}$
and by $t_i^{\prime\, 2}\br(a_i\to b\anti b)$ for $a_{1,2}$.

\begin{table}[h!]
\caption{
$t \bar{t} h \to t\bar{t}b\bar{b}$: $L=100\fbi$, $K_S=K_B=1$,
quoted for $\br(h\to b\bar b)=1$ \label{chan3}}
\smallskip
\begin{center}
\footnotesize
\begin{tabular} {|l||l|l|l|l|l|l|l|l|} 
\hline
$m$ [GeV] & 80  & 90  & 100 & 110  & 120  & 130  & 140  & 150\\
\hline
$S/\sqrt{B}$   & 17.9  & 15.0  & 14.1 & 12.3 & 12.7  & 13.7  & 11.3  & 10.6 \\
\hline
\end{tabular}
\end{center}
\end{table}

\item{Channel 4):} For $b\anti b h/a\to b\anti b \tauptaum$
we have employed the experimental studies presented in 
\cite{unknown:1999fr} (as contained in the $L=100~\fbi$ curve of
fig.~19-62 and also using information in 
tables~19.35/36). These results were repeated (for the mass range below
$500\gev$ where we employ them) in the Les Houches workshop study of
\cite{Cavalli:2002vs} fig.~E.15.  
The estimation of the statistical
significances using fig.~19-62 of \cite{unknown:1999fr} for this channel
requires considerable discussion. 

Figure 19-62 
 gives the $5 \sigma$ contours in the $\tan\beta$ - $m_A$ plane of the
MSSM. The critical issue is what fraction of these $5\sigma$ signals derives
from $gg\to H / gg\to A$ production and what fraction from associated $b\anti b
H / b\anti
bA$ production, and how each of the $gg$ fusion and $b\anti b$ associated
production processes are divided up between $H$ and $A$. For the former, we
turn to table 19.35 of ref. \cite{unknown:1999fr}. There, we see that
it is for cuts 
designed to single out the associated production processes that large
statistical significance can be achieved and that such cuts provide 90\% of the
net statistical significance of $N_{SD}=8.9$ (3.9 for $gg$ fusion cuts combined
in quadrature with 8.0 for $b\anti bH+b\anti bA$ associated production cuts)
for $m_A=150\gev$ and $L=30\fbi$. (For the associated production cuts, the
table of \cite{unknown:1999fr}
shows that the contribution of the $gg$ fusion processes to the signal is
very small.) The percentage of $N_{SD}$ deriving from $gg$-fusion cuts is even
smaller at high $m_A$. For $m_H\sim m_A\in
[100\gev,500\gev]$, a conservative choice is then that 90\% of the statistical significance
along the contours of fig. 19-62 comes from the associated production cut
analysis. With this choice, the
$5\sigma$ contour at $L=100\fbi$ from fig. 19-62 of ref. \cite{unknown:1999fr} corresponds
to a $4.5\sigma$ contour for associated $b\anti b H+b\anti bA$ production
alone. Since the values of $\tanb$ along this contour are large, we can
separate the $H$ and $A$ signals from one another by using the following
properties of the MSSM within which fig. 19-62 of ref. \cite{unknown:1999fr} was
generated: (a) $\br(H\to \tauptaum)\sim \br(A\to\tauptaum)\sim 0.09$; 
(b) the $b\anti b A$ and $b\anti b H$ couplings are very nearly equal and scale
as $\tan\beta$; and (c) $m_A\sim m_H$ within the $\tauptaum$ mass
resolution. As a result, the net signal rate along this contour is
approximately twice that for $b\anti b A$ or $b\anti b H$ alone. Thus,
$N_{SD}=2.25$ would be achieved for $b\anti b A$ or $b\anti b H$ along this
contour were $m_A$ and $m_H$ widely separated.
Defining the
value of $\tanb$ as a function of $m_A$ shown by the $100\fbi$ curve of fig.
19-62 in ref. \cite{unknown:1999fr} as $\tanb_{2.25}(m_A)$, we compute 
\beq
N_{SD}(\tanb=1,m)=2.25 \left[\frac{1}{\tanb_{2.25}(m)}\right]^2
\eeq
These are the numbers tabulated in table~\ref{chan4} below (where, for
convenience, we include an extra factor of 100).

The above procedure is conservative in that it assumes no contribution to the
$\tauptaum$ channel $N_{SD}$ from the $gg$ fusion processes. 
We have not attempted to include the latter production process, since 
the $\tauptaum$ mode is only useful in finding $5\sigma$ contours 
when the $b\anti b$ Higgs coupling is highly enhanced, in which case
the $gg$ fusion process will make a relatively very small contribution.

Finally, the results of ref.~\cite{unknown:1999fr}
assumed $K_S=K_B=1$ and assumed the MSSM mass-independent value for the
branching ratio, $\br(H,A\to \tauptaum)=0.09$.
Putting all this together,
the results of table~\ref{chan4} must be rescaled by 
the factor $0.01 b_i^2\br(h_i\to \tauptaum)/0.09$ for $h_{1,2,3}$
and by $0.01 b_i^{\prime\, 2}\br(a_i\to \tauptaum)/0.09$ for $a_{1,2}$.
The $L=300\fbi$ results are obtained
by scaling the results so obtained by
$\left[\frac{300\fbi}{100\fbi}\right]^{1/2}$.
\begin{table}[h!]
\caption{
 $b\bar{b} h~or~b\bar b a \to b\anti b\tau\bar{\tau}$ at $g_{b\anti b
   h,a}=g_{b\anti b H,A}(\tanb=1)$: $L=100\fbi$, $K_S=K_B=1$ \label{chan4}}
\begin{center}
\begin{tabular} {|l||l|l|l|l|l|l|l|} 
\hline
$m$ [GeV] & 100 & 110 & 120 & 130 & 140 & 150 & 200 \\
\hline
$S/\sqrt{B}$  (x$10^2$)  & 3.7 & 4.2 & 4.4 & 4.5 & 4.7 & 4.6
& 3.1 \\
\hline
\hline
$m$ [GeV] & 250 & 300 & 350 & 400 & 450 & 500 & \\
\hline
$S/\sqrt{B}$  (x$10^2$)  & 2.1 & 1.3 & 1.0 & 0.8 & 0.7 & 0.6 &
\\
\hline
\end{tabular}
\end{center}
\end{table}

\item{Channel 5:} For $gg\to h\to ZZ^{(*)}\to 4\ell,\ell\ell\nu\nu$,
  we employ the CMS ``no K-factors'', $L=100\fbi$ plot  supplied to us by
  R. Kinnunen. For Higgs mass below $500\gev$, only the $4\ell$ mode
is present on the plot.  For masses from $500\gev$ up to $1\tev$,
the tabulated numbers were obtained by combining in quadrature  the
plotted results for the $4\ell$ and $2\ell 2\nu$ modes.
See also, \cite{Kinnunen:1997}. (The
CMS $L=30\fbi$ results appear in fig.~1  of
  \cite{Kinnunen:2002} and figs.~12 and 13 of \cite{Abdullin:2003}.)  
The results quoted in table~\ref{chan5} assume $K_S=K_B=1$. 
The tabulated values are to be multiplied by $R_i^2 \br(h_i\to
ZZ^{(*)})/\br(\hsm\to ZZ^{(*)})$ for $h_{1,2,3}$.  We assume no
contribution from this mode to $a_{1,2}$ corresponding to the absence
of tree-level $ZZa_i$ couplings.  For $L=300\fbi$, we scale by the
factor $\left[\frac{300\fbi}{100\fbi}\right]^{1/2}$.
\begin{table}[h!]
\caption{
 $gg \to h \to ZZ^{(*)} \to 4\ell, \ell\ell\nu\nu$: $L=100\fbi$,
 $K_S=K_B=1$ \label{chan5}}
\smallskip
\begin{center}
\begin{tabular} {|l||l|l|l|l|l|l|l|l|l|l|l|} 
\hline
$m$ [GeV]  & 100 & 120 & 130 & 140  & 150 & 160  & 170 & 180 &
190 & 200 \\
\hline
$S/\sqrt{B}$   & 2.7 & 5.3 & 13.2 & 22.1  & 27.8 & 9.4  & 5.5 & 20.7 &
25.1 & 26.1 \\
\hline
\hline
$m$ [GeV]  & 250 & 275  & 350 & 400 & 500 & 600 & 700 & 800 & 1000 & 
\\
\hline
$S/\sqrt{B}$  & 21.6 & 17.6  & 22.7 & 21.6 & 21.5
& 17.1 & 13.6 & 11.1 & 9.3 & \\
\hline
\end{tabular}
\end{center}
\end{table}

\item{Channel 6):} For $gg\to h \to WW^{(*)}\to \ell\ell\nu\nu,\ell\nu
  jj$, we again employ the CMS $K_S=K_B=1$, $L=100\fbi$ plot supplied to us by
  R. Kinnunen. The $\ell\ell\nu\nu$ signal is the only
one present for $[120\gev,250\gev]$. At mass=$300\gev$, both
$\ell\ell\nu\nu$ and $\ell\ell jj$ are present, and we combine them in
quadrature.  For the masses of $600\gev$ and $800\gev$, only the
$\ell\ell jj$ signal is present.  The results that we obtain
in this way from the CMS plot areas tabulated in table~\ref{chan6} below. 
(The
CMS $L=30\fbi$ results appear in fig.~1  of
  \cite{Kinnunen:2002} and figs.~12 and 13 of \cite{Abdullin:2003}.)
  For NMSSM Higgs statistical significances at $L=100\fbi$, we
  multiply the values in the table 
  by $g_i^2 \br(h_i\to WW^{(*)})/\br(\hsm\to WW^{(*)})$ for
  the $h_{1,2,3}$. This channel is absent at tree-level for the
  $a_{1,2}$. In going to $L=300\fbi$, results obtained in this way
were multiplied by $\left[\frac{300\fbi}{100\fbi}\right]^{1/2}$.
\begin{table}[h!]
\caption{ $gg \to h \to WW^{(*)} \to \ell\ell\nu\nu, \ell\nu jj$:
  $L=100\fbi$, $K_S=K_B=1$ \label{chan6}}
\smallskip
\begin{center}
\begin{tabular} {|l||l|l|l|l|l|l|l|l|} 
\hline
$m$ [GeV] & 120 & 130 & 140 & 150 & 160 & 170 & 180  \\
\hline
$S/\sqrt{B}$  & 5.1 & 9.8 & 17.8 & 21.9 & 47.0 & 34.4 &
24.1  \\
\hline
\hline
$m$ [GeV]& 190 & 200  & 250 & 300 & 600 & 800 &  \\
\hline
$S/\sqrt{B}$& 19.5  & 16.9  & 7.9 & 19.4  & 14.2 &
11.3 &  \\
\hline
\end{tabular}
\end{center}
\end{table}

\item{Channel 7):} For the $W^+W^-\to h \to \tauptaum$ channel, we
  employed the table~10, $L=30\fbi$ ATLAS results of \cite{asai:2003}, rescaled
  to $L=100\fbi$ by the factor of $\left[\frac{100\fbi}{30\fbi}\right]^{1/2}$.
  The results of this rescaling are given in table~\ref{chan7} below. The
  tabulated values are multiplied by
  $R_i^2\br(h_i\to\tauptaum)/\br(\hsm\to\tauptaum)$ for the NMSSM $h_{1,2,3}$.
  This mode is not present (at tree-level) for the $a_{1,2}$. In going to
  $L=300\fbi$, we multiplied the results so obtained by the somewhat
  conservative factor of $\left[\frac{300\fbi}{100\fbi}\right]^{1/4}$.
\begin{table}[h!]\caption{$WW \to h \to \tauptaum$: $L=100\fbi$,
    $K_S=K_B=1$ \label{chan7}}
\smallskip
\begin{center}
\begin{tabular} {|l||l|l|l|l|l|l|l|l|l|l|l|l|l|l|l|} 
\hline
$m$ [GeV]  & 110  & 120  & 130 & 140  &
150 \\
\hline
$S/\sqrt{B}$  & 6.7  & 10.4 & 10.4  &
8.7 & 4.4  \\
\hline
\end{tabular}
\end{center}
\end{table}

\item{Channel 8):} For $WW \to h \to WW$, we employed the results in
  table~7 of \cite{asai:2003} in the last row labeled ``combined
  statistical significance''.  These results were those obtained for
  $L=10\fbi$. Since the main final state contributors to the
statistical significances given for $L=10\fbi$ were the $WW\to ee$,
$\mu\mu$ and $e\mu$ final states, we felt that these results could
safely be scaled up to $L=100\fbi$ using the factor
$\left[\frac{100\fbi}{10\fbi}\right]^{1/2}$. The results of this scaling are tabulated in
table~\ref{chan8}. In addition, there was a specialized neural net
analysis for the limited mass range of $[115\gev,130\gev]$ \cite{unknownatlas:2003}.
The results corresponding to $L=100\fbi$ from table~5 of this analysis are given
in the parentheses in table~\ref{chan8}. In the $[115\gev,130\gev]$ mass range, we have employed
the (stronger) neural net result. Entries in
table~\ref{chan8} are to be multiplied by $R_i^2\br(h_i\to
W^+W^-)/\br(\hsm\to W^+W^-)$ for the $h_{1,2,3}$. The process is
absent at tree-level for the $a_{1,2}$.
In going to $L=300\fbi$, we have been somewhat
conservative and scaled the results so obtained by the factor
$\left[\frac{300\fbi}{100\fbi}\right]^{1/4}$. 

\begin{table}[h!]\caption{$WW \to h \to WW$: $L=100\fbi$, $K_S=K_B=1$ \label{chan8}}
\smallskip
\begin{center}
\begin{tabular} {|l||l|l|l|l|l|l|l|l|l|l|} 
\hline
$m$ [GeV]  & 110 & 115 & 120 & 125 & 130  & 140  \\
\hline
$S/\sqrt{B}$   & 2.5 & (5.6) & 6.6 (9.7) & 15.7 & 13.9
(20.5)  & 18.6 \\
\hline
\hline
$m$ [GeV] & 150  & 160  & 170  & 180  & 190 & \\
\hline
$S/\sqrt{B}$ & 26.5  & 34.8  & 34.8 & 27.8  &
21.5 & \\
\hline
\end{tabular}
\end{center}
\end{table}

\item{Channel 9):} For invisibly decaying Higgs bosons, 
there are two experimental studies. The first is that of 
\cite{nikmaz:2002} covering the Higgs mass range $[100\gev,140\gev]$.
This study was recently extended to a larger
Higgs mass range in \cite{Abdullin:2003}. 
Both studies were performed for $L=10\fbi$.  Since we are uncertain
that these results can be easily employed at higher $L$, our
program currently assumes that only $L=10\fbi$ of data is accumulated
for the $WW\to h\to {invisible}$ mode. 
The appropriate procedure
for the results quoted in \cite{Abdullin:2003} as
based on \cite{Cavalli:2002vs} is as follows.  The raw
$S/\sqrt B$ of this latter reference
agrees (at Higgs mass $=120\gev$) 
with that in \cite{nikmaz:2002}. However, \cite{Abdullin:2003} 
includes a systematic 3\% uncertainty in the background
and computes $S/\sqrt{B+(0.03 B)^2}$ to obtain
the 95\% CL limits of their fig.~25.  This we believe is 
the more reliable way of estimating the significance of
the signal given the amorphous nature of the backgrounds.
In table~\ref{chan9}, we give the significances
after accounting for the background systematic uncertainty.
These are 
extracted from  fig.~25 of \cite{Abdullin:2003} using
the formula $S/\sqrt{B+(0.03 B)^2}=1.96/\xi^2$,
where $\xi^2=\br(h\to invisible)\frac{\sigma(WW\to h)}{
  \sigma(WW\to\hsm)}$ is the quantity plotted.

\begin{table}[h!]\caption{$WW \to h \to invisible$: $L=10\fbi$,
    $K_S=K_B=1$, $\br(h\to invisible)=1$, SM $WWh$ coupling \label{chan9}}
\smallskip
\begin{center}
\begin{tabular} {|l||l|l|l|l|l|l|l|l|} 
\hline
$m$ [GeV] & 120 & 150 & 200 & 250 & 300 & 350 & 400 \\
\hline
$\frac{S}{\sqrt{B+(0.03B)^2}}$  & 15 & 14 & 13 & 11 & 10 & 8 & 7 \\
\hline
\end{tabular}
\end{center}
\end{table}

In the above formulae,
\bea
\br(h_1\to invisible)&=& \br(h_1\to \cnone\cnone)+\br(h_1\to
a_1a_1)[\br(a_1\to\cnone\cnone)]^2 \\
\br(h_2\to invisible)&=& \br(h_2\to \cnone\cnone)\nn\\
&& +\br(h_2\to a_1a_1)[\br(a_1\to\cnone\cnone)]^2\nn\\
&&+\br(h_2\to
h_1h_1)[\br(h_1\to invisible)]^2 \\
\br(h_3\to invisible)&=& \br(h_3\to \cnone\cnone)\nn\\
&&+\br(h_3\to a_1a_1)[\br(a_1\to\cnone\cnone)]^2\nn\\
&&+\br(h_3\to h_1h_1)[\br(h_1\to invisible)]^2\nn\\
&& +\br(h_3\to h_1h_2)\br(h_1\to invisible)\br(h_2\to invisible)\nn\\
&&+\br(h_3\to h_2 h_2)[\br(h_2\to invisible)]^2\,.
\eea

\eed

\newpage

\bibliography{nmssmlhchiggs}

\end{document}